\documentclass[sigplan,twocolumn]{acmart}
\renewcommand\footnotetextcopyrightpermission[1]{}
\usepackage{graphicx} 
\usepackage{algorithm}
\usepackage{algpseudocode}
\usepackage{subcaption}
\usepackage{booktabs}

\newcommand{\NAME}{\textsf{RServe}}
\newcommand{\CP}{\textsf{Sarathi-Serve}}

\newcommand{\xwz}[2]{\textcolor{black}{#1}} 

\title{\NAME: Overlapping Encoding and Prefill for Efficient LMM Inference}

\author{Tianyu Guo}
\email{guoty9@mail2.sysu.edu.cn}
\orcid{0009-0005-2979-4486}
\affiliation{
    \institution{CSE, Sun Yat-sen University} 
    \city{Guangzhou}
    \country{China}
}

\author{Tianming Xu}
\email{zhuran@xiaohongshu.com}
\affiliation{
    \institution{Rednote} 
    \city{Beijing}
    \country{China}
}

\author{Xianjie Chen}
\email{chenxj275@mail2.sysu.edu.cn}
\affiliation{
    \institution{CSE, Sun Yat-sen University} 
    \city{Guangzhou}
    \country{China}
}

\author{Junru Chen}
\email{chenjr97@mail2.sysu.edu.cn}
\affiliation{
    \institution{CSE, Sun Yat-sen University} 
    \city{Guangzhou}
    \country{China}
}

\author{Nong Xiao}
\email{xiaon6@mail.sysu.edu.cn}
\orcid{0000-0002-2166-977X}
\affiliation{
    \institution{CSE, Sun Yat-sen University} 
    \city{Guangzhou}
    \country{China}
}

\author{Xianwei Zhang}
\email{zhangxw79@mail.sysu.edu.cn}
\orcid{0000-0003-3507-4299}
\affiliation{
    \institution{CSE, Sun Yat-sen University} 
    \city{Guangzhou}
    \country{China}
}

\begin{abstract}
\xwz{L}{Current l}arge multimodal models (LMMs) typically employ an encoding module to \xwz{transform}{convert} multimodal data \xwz{inputs into}{into input} embeddings, which are then fed to language models for \xwz{further}{} processing. 
However, efficiently serving LMMs remains highly challenging due to the inherent complexity of their inference pipelines.
Traditional serving engines co-locate the encoding module and the language model, leading to significant resource interference and tight data dependency. 
Recent studies have \xwz{alleviated}{addressed} this issue by disaggregating the encoding module from \xwz{the model}{language models}, \xwz{following a design style of prefill-decode disaggregation}{adopting architectures similar to the prefill-decode disaggregated design}. 
Nevertheless, these approaches fail to fully \xwz{exploit parallelism}{unlock the parallel processing potential} both within \xwz{individual requests}{a single request} (intra-request) and across multiple requests (inter-request).

To \xwz{overcome}{tackle} the limitation, we propose \NAME, an LMM inference system \xwz{that}{designed to} efficiently orchestrate\xwz{s}{} intra- and inter-request pipelines. 
\xwz{\NAME~is designed to reduce}{Its core objectives are to achieve} low latency and maximize parallelism \xwz{at both intra- and inter-request granularities}{both within individual requests and between multiple requests}.
Built on the disaggregated architecture of the encoding module and language model, \NAME~adopts a fine-grained scheduling method that \xwz{overlaps}{enables overlapping between} multimodal encoding \xwz{with}{and} the forward computation of the language model within a single request. 
For inter-request pipeline\xwz{}{optimization}, \NAME~leverages schedulable tokens and token budgets to balance computational loads across micro-batches. 
Combined with chunked prefill, this \xwz{enables}{forms} a novel scheduling \xwz{strategy}{approach} that \xwz{coordinates the}{enables coordinated} execution of intra-\xwz{}{request} and inter-request pipelines.
Experimental evaluations on representative LMMs show that \NAME~achieves substantial latency reduction of up to 66\% while improving throughput by up to 109\%\xwz{, significantly outperforming existing serving approaches}{}.
\end{abstract}



\begin{document}

\fancyhead{}

\maketitle

\section{INTRODUCTION}

As large language models (LLMs) \cite{DBLP:journals/corr/abs-2507-20534,DBLP:journals/corr/abs-2505-09388,DBLP:journals/corr/abs-2508-10925} and large multimodal models (LMMs) \cite{Qwen2_5_VL,DBLP:journals/corr/abs-2412-10302,Llama4,InternVL,LLaVA-OneVision,NVLM} find widespread applications across diverse fields \cite{DBLP:conf/aaai/ZhangLLXYLWCLQY25,DBLP:conf/icse/Wang0ZFZ0025,DBLP:conf/aaai/AnandPKNNJS25,DBLP:conf/aaai/KangWJWH025,DBLP:conf/aaai/YeGZ0W25,DBLP:conf/aaai/00010BZ0025,DBLP:conf/aaai/SouzaCF25,EFIM,DBLP:conf/aaai/00020SWZZY0025,DBLP:conf/aaai/GaoQWWCL25,DBLP:conf/aaai/WuZCXJWLHQF25}, efficient \xwz{inference}{} serving has become a critical research focus in both industry and academia.
While extensive studies have investigated the serving processes of LLMs, multimodal models present \xwz{different challenges}{a different challenge}. Their unique inference pipelines introduce additional complexities, shifting the problem space and requiring new approaches beyond those designed for text-only models.
\xwz{Model}{LLM} inference consists of two core procedures: prefill and decode \cite{Orca}.
The prefill stage computes the key-value (KV) cache \cite{vllm} for the entire input prompt and generates the first output token, which is compute-bound \cite{Sarathi-serve}.
The decode stage generates subsequent tokens in an autoregressive manner, which is memory-bound.
To \xwz{cope with}{address} the distinct computational characteristics of these two stages, recent \xwz{researches have}{research has} proposed a prefill-decode disaggregated architecture \cite{Splitwise, DistServe, Mooncake}. 
In \xwz{such architecture}{which}, prefill and decode operations are allocated to separate nodes, with communication between nodes \xwz{being}{} enabled via KV cache transmission \cite{Mooncake}. 

\begin{figure}[t]
    \centering
    \includegraphics[width=\linewidth]{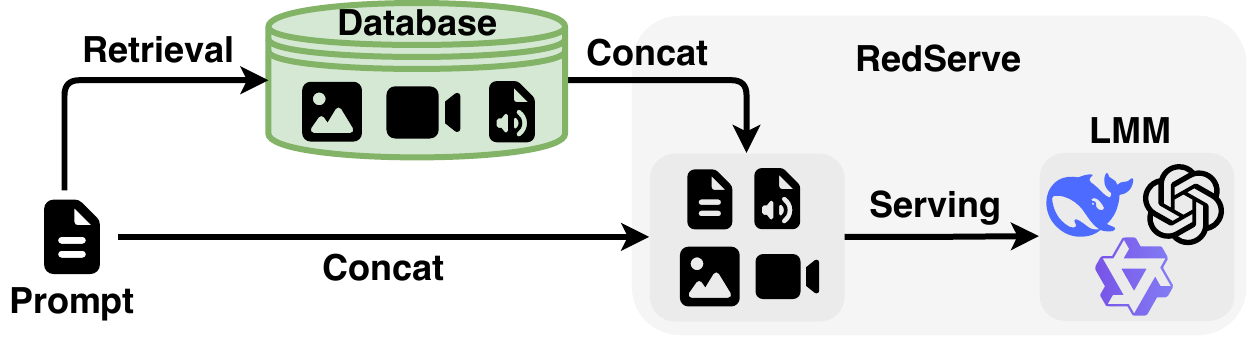}
    \caption{\xwz{LMM serving}{Serving LMMs} allows prompts to incorporate increasingly rich and diverse multimodal data.}
    \label{fig:scene}
\end{figure}

\begin{figure}[t]
    \centering
    \includegraphics[width=\linewidth]{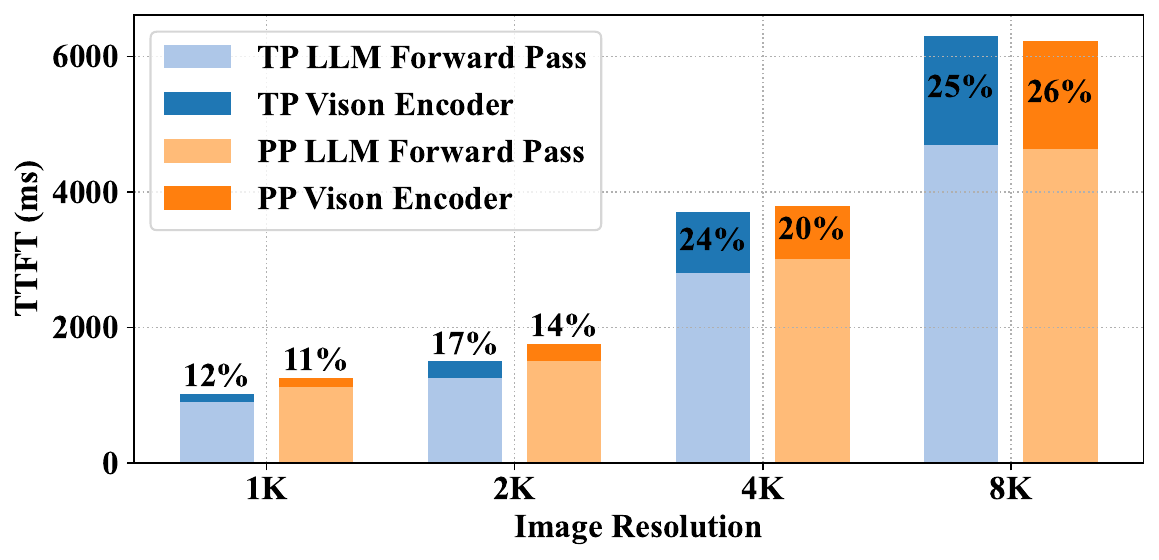}
    \caption{Single multimodal request (with two images) latency breakdown for tensor parallelism and chunked pipeline parallelism (4$\times$H100) as image resolution increases.
    The numbers on the bar represent the proportion of total latency accounted for by the multimodal encoding time.}
    \label{fig:intro}
\end{figure}

Compared with pure language models, multimodal model inference often relies on substantially richer prompt inputs, as shown in Figure \ref{fig:scene}.
Current LMMs rely on an additional encoder to produce multimodal\footnote{In this paper, multimodal refers to modalities other than text.} input embeddings that are compatible with those used in traditional LLMs.
Nonetheless, LMMs face another challenge: encoding all multimodal data within a single request introduces significant latency (the proportion can reach up to 26\%\xwz{, as}{} shown in Figure \ref{fig:intro}), which interferes with the forward computation LLMs.
Recent studies \cite{ModServe,EPD} have attempted to address this interference by disaggregating the multimodal encoding module from the LLMs. Unfortunately, \xwz{such}{this} disaggregation fails to fully leverage intra-request parallelism between the multimodal encoding process and LLM forward computation.
Current inference systems typically treat the encoder and prefill stages as strictly sequential \cite{vllm,SGLang}, where prefill for a request only starts after all its multimodal information has been fully encoded. 
By leveraging chunked prefill, however, a portion of the prepared embeddings can be processed in advance. 
This allows the multimodal encoding and prefill operations within a request to overlap, effectively \xwz{lowering}{reducing} the request’s end-to-end latency.

As model sizes continue to grow, distributed deployment of LLMs or LMMs has become mainstream. 
Among the most widely used methods are tensor parallelism and pipeline parallelism. 
Tensor parallelism is typically employed for intra-node \xwz{}{parallelism}with high-bandwidth interconnects and can effectively reduce inference latency. 
Pipeline parallelism, on the other hand, is generally used for inter-node \xwz{case}{parallelism} with limited bandwidth and can improve inference throughput. 
The recently proposed chunked pipeline parallelism (CPP) \cite{Mooncake} changes this landscape, enabling pipeline parallelism to achieve latency reductions comparable to tensor parallelism. 
Specifically, CPP splits the entire input embedding into multiple chunks and pipelines the prefill computation in the original input order, allowing different chunks of a single request to be processed simultaneously within the pipeline. 
As shown in Figure \ref{fig:intro}, pipeline parallelism and tensor parallelism can maintain \xwz{comparable}{similar} inference latency for a single multimodal request.

To enable interleaved and overlapped execution, we propose \NAME, a LMM inference system that orchestrates intra-request and inter-request pipelines with full parallelism. 
For requests with rich multimodal inputs (i.e., lots of images), \NAME~overlaps the multimodal encoding process with prefill execution, constructing intra-request pipeline.
To realize this, \NAME~categorizes input embeddings into two types: ready embeddings and not-ready \xwz{ones}{embeddings}. 
Ready embeddings comprise text embeddings and already encoded multimodal \xwz{ones}{embeddings}, whereas not-ready embeddings refer to those that have not been processed by the encoding module. 
\NAME~encodes multimodal data sequentially from left to right at a fine granularity, allowing the LLM to initiate prefill execution as soon as partial embeddings are produced.
To maintain high throughput \xwz{and}{while achieving} low latency, \NAME~further batches distinct requests for execution and employs schedulable tokens to balance the computational load across individual micro-batches, building an inter-request pipeline.
\NAME's intra-request pipeline is an optimization that is independent of the parallelism method, whereas inter-request pipeline combines intra-request pipeline with pipeline parallelism.

The contributions of this paper are:
\begin{itemize}
    \item We highlight intra-request parallelism between the multimodal encoding and LLM forward pass has not been fully utilized.
    \item We propose \NAME, an efficient LMM inference system that orchestrates intra- and inter-request pipel-ine to reduce latency while maintaining high throughput.
    \item Experimental results on representative LMMs demonstrate that \NAME~reduces latency by as much as 66\% and improves throughput by up to 109\%.
\end{itemize}

\section{BACKGROUND AND MOTIVATION}

\subsection{Model Inference Procedure}

\subsubsection{LLM Inference Procedure}

\begin{figure}[ht]
    \centering
    \includegraphics[width=\linewidth]{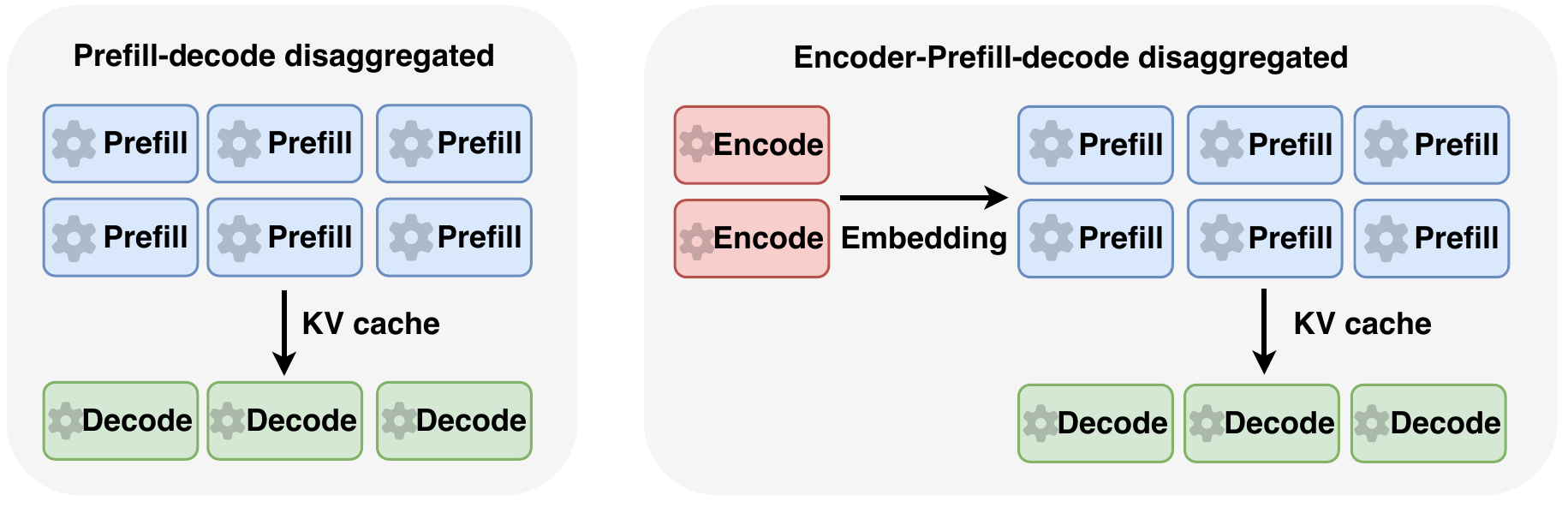}
    \caption{The inference paradigms of LLMs and LMMs differ: LLMs employ a prefill-decode disaggregated architecture, whereas LMMs utilize an encoder-prefill-decode (EPD) disaggregated architecture.}
    \label{fig:pd-epd}
\end{figure}

In LLM inference, token generation proceeds in an autoregressive manner \cite{Transformer}, where each token is conditioned on all previously generated tokens. 
To reduce redundant computation, modern serving systems leverage the KV cache \cite{Orca, vllm, SGLang}, which stores intermediate states required for decoding. 
Based on computational characteristics, the inference pipeline can be divided into two phases\xwz{~of}{:} prefill and decode. 
The prefill phase processes the entire input prompt, constructs the KV cache, and produces the first output token, typically leading to high GPU utilization. 
In contrast, the decode phase generates subsequent tokens by reusing the KV cache; GPU utilization in this phase is relatively low, and batching across multiple requests is \xwz{thus}{} commonly employed to improve efficiency.

To alleviate the latency bottleneck introduced by prefill, recent work \cite{Sarathi-serve} proposes chunked prefill, which partitions the prefill computation into smaller segments and interleaves their execution with batched decoding.
While this strategy reduces the delay of decode, it cannot fully eliminate the resource interference between prefill and decode. 
To address this issue, a prefill–decode disaggregated \cite{DistServe, Splitwise, Mooncake} architecture (as shown in Figure \ref{fig:pd-epd}) has been \xwz{proposed}{introduced}, in which prefill and decode operations are dispatched to separate nodes, with the KV cache transmitted across them to enable efficient collaboration \cite{Mooncake}.

\subsubsection{LMM Inference Procedure}

\begin{figure}[t]
    \centering
    \includegraphics[width=\linewidth]{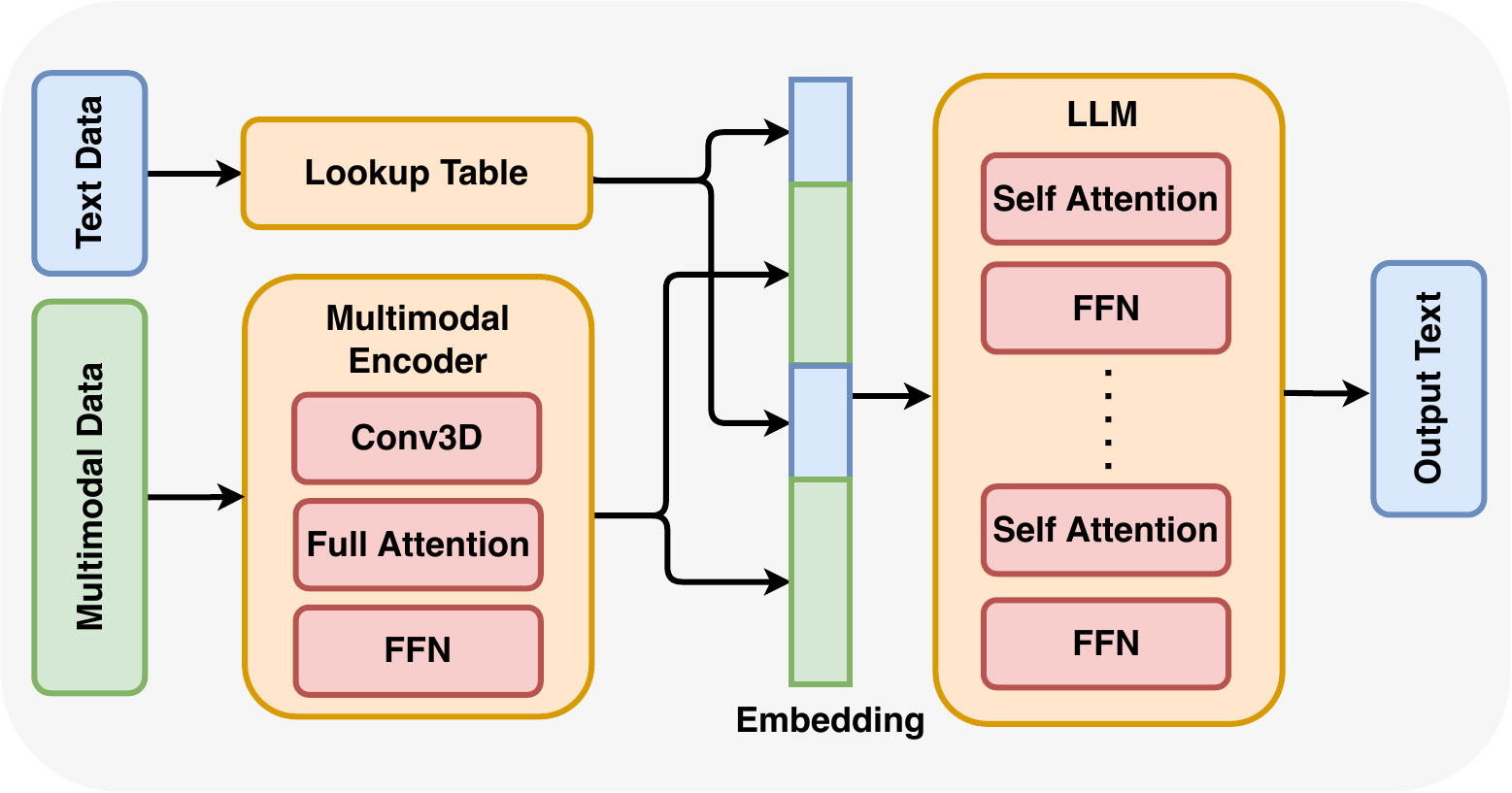}
    \caption{LMM inference diagram. Text data and multimodal data are encoded through separate pathways to generate embeddings, which are then combined and fed into the LLM to generate text. In general, the encoding overhead for multimodal data is usually higher \xwz{than that of text}{}.}
    \label{fig:lmm}
\end{figure}

\begin{figure}[t]
    \centering
    \includegraphics[width=\linewidth]{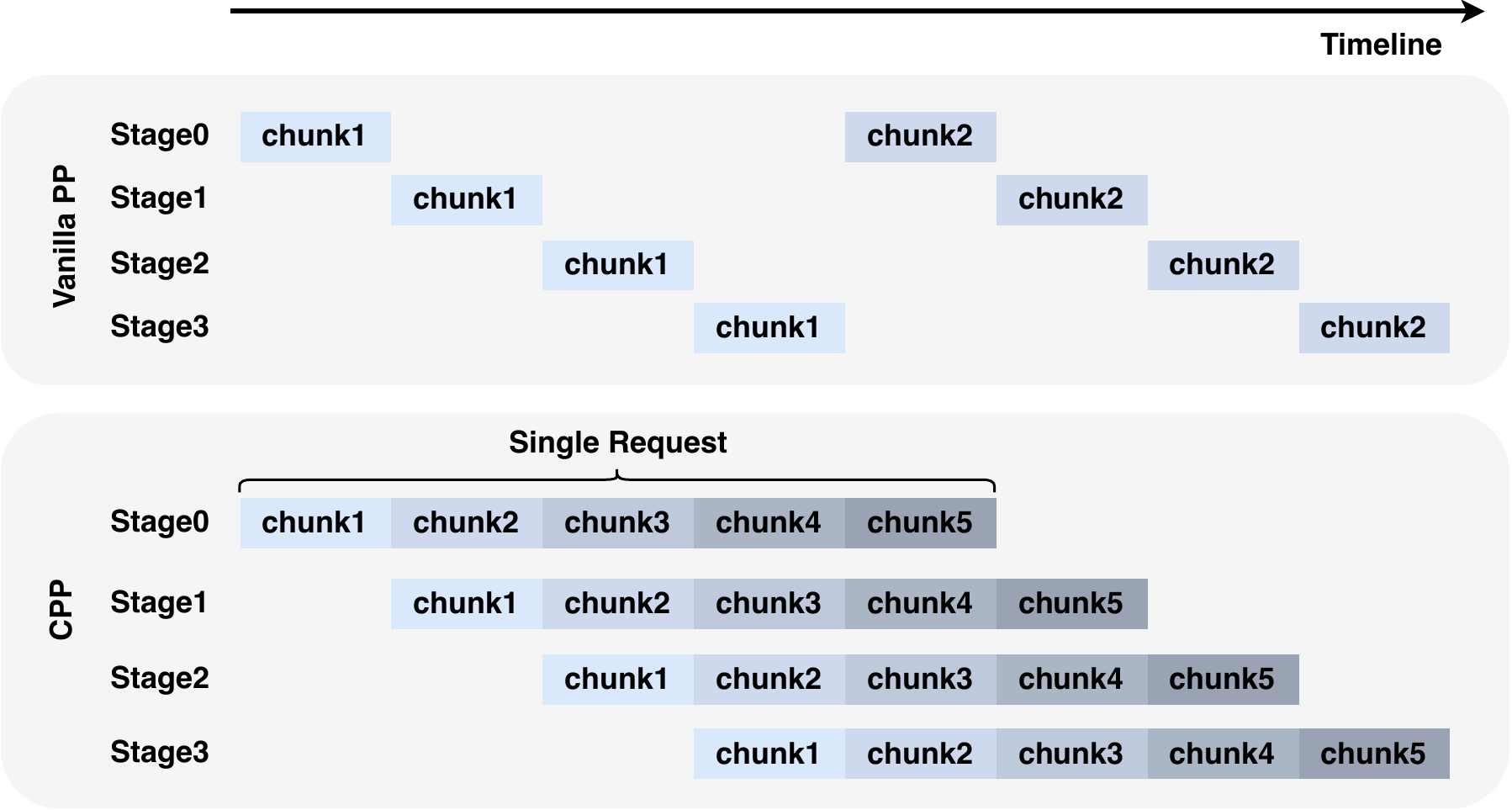}
    \caption{Comparison between vanilla PP and CPP. Vanilla PP starts the prefill computation of next chunk after the finish of previous chunk. CPP begins the prefill computation of next chunk once the finish of previous \xwz{one}{chunk} at each stage.}
    \label{fig:cpp}
\end{figure}

\begin{figure*}[ht]
    \centering
    \includegraphics[width=.9\linewidth]{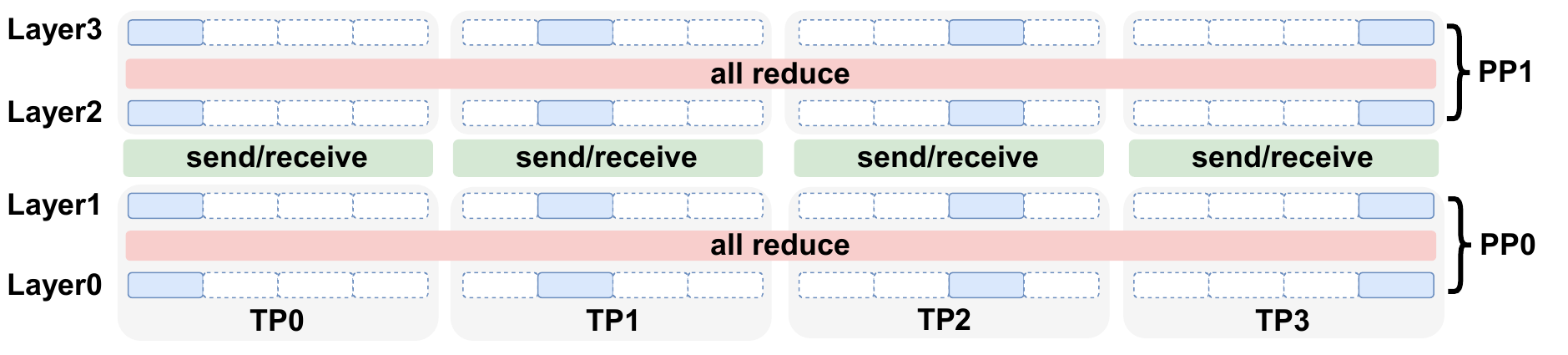}
    \caption{An illustration of combining TP ($\times$4) and PP ($\times$2). TP partitions the model parameters within each layer, while PP partitions them across layers. TP \xwz{necessitates}{requires} synchronous communication among the ranks within a TP group, whereas PP only requires asynchronous point-to-point communication at the boundaries of the layer partitions.}
    \label{fig:pp_tp}
\end{figure*}

As illustrated in Figure \ref{fig:lmm}, LMMs initially employ a multimodal encoder to convert multimodal inputs into embeddings \cite{ModServe, EPD}. 
This encoder typically comprises components such as 3D convolutional layers, attention mechanisms, and feed-forward networks (FFNs), which are designed to capture both spatial and temporal dependencies across different data modalities. 
In comparison, textual data require\xwz{s}{} only a vocabulary lookup to obtain token embeddings, which is computationally trivial. 
Consequently, the encoding of multimodal inputs introduces a substantial computational overhead, particularly when processing high-resolution images, long video sequences, \xwz{and}{or} complex audio signals.

The embeddings derived from multimodal and textual inputs are subsequently integrated and passed into LLM for further reasoning and generation tasks. 
As the volume and complexity of multimodal data increase, the encoding stage becomes a critical bottleneck\xwz{}{ in LMM inference}, often making a great contribution to the overall \xwz{inference}{} latency. 
This bottleneck has significant implications for real-time applications, such as interactive multimodal assistants or autonomous systems, where both high accuracy and low latency are essential.
Addressing this challenge requires careful optimization of coordinating encoder and LLM computations.

Recent research\xwz{es}{} \cite{ModServe,EPD} suggest that co-locating encoding and prefill operations can intensify interference between them, as each must wait for the other to complete. To address this issue, a recent study proposes an encoder–prefill-decode (EPD) disaggregated architecture (Figure \ref{fig:pd-epd}), where the encoder and prefill computations are executed on separate devices or nodes. In this design, the encoder worker is dedicated solely to multimodal data encoding and transmits the result\xwz{ed}{ing} embeddings to the prefill worker. Once the prefill worker receives these embeddings, it can immediately begin prefill computation. This separation eliminates mutual interference between encoding and prefill operations.

\subsection{\xwz{Model Inference Parallelism}{Parallelism in LLM Inference}}
The parallelization strategies for large models can be broadly categorized into data parallelism and model parallelism. 
Data parallelism distributes incoming requests across multiple inference instances, while model parallelism partitions the model parameters across different GPUs, enabling collaborative inference among them. 
Model parallelism can be further divided into tensor parallelism and pipeline parallelism.

Tensor parallelism implements intra-layer \xwz{concurrency}{parallelism} by dividing individual operations within a single layer across multiple devices. 
This approach is well known for its ability to significantly reduce the latency of single-request inference, as computations within a layer are executed concurrently. 
In contrast, pipeline parallelism exploits inter-layer \xwz{concurrency}{parallelism}, assigning consecutive layers of the model to different devices. 
By processing multiple requests simultaneously, it primarily improves overall throughput. 
Traditionally, pipeline parallelism has been regarded as less effective for lowering forward-pass latency due to the sequential dependency between layers. 
However, this limitation can be mitigated through CPP \cite{Mooncake}, which partitions a single input into smaller micro-batches and feeds them through the pipeline in a staggered manner. 
By overlapping computation across layers for the same request, CPP effectively leverages intra-request parallelism to achieve substantial latency reduction, while still benefiting from the throughput advantages of standard pipeline parallelism.

\subsubsection{Chunked Pipeline Parallelism}

Current model serving systems use chunked prefill to process ultra-long context. 
Specifically, the whole prompt is split into multiple chunks to be processed one by one.
Since the prefill computation of one individual chunk only relies on the preceding \xwz{ones}{chunks}, we can pipeline the computation of these chunks and overlap the execution of different chunks.
Once the previous chunks have finished at a stage, the next chunk can leverage the KV cache of previous \xwz{ones}{chunks} and begin the prefill computation.
In this way, CPP can greatly lower the latency of prefill computation. 
Figure \ref{fig:cpp} shows a comparison of vanilla PP and CPP.
In vanilla PP, the prefill computation of \xwz{}{the}next chunk can only begin after the previous chunk has fully completed all stages. 
In contrast, CPP allows the prefill operations of the next chunk to start as soon as the previous chunk has finished a given stage.

\subsection{Intra- and Inter-request Parallelism in LMM}

\begin{figure}[ht]
    \centering
    \includegraphics[width=\linewidth]{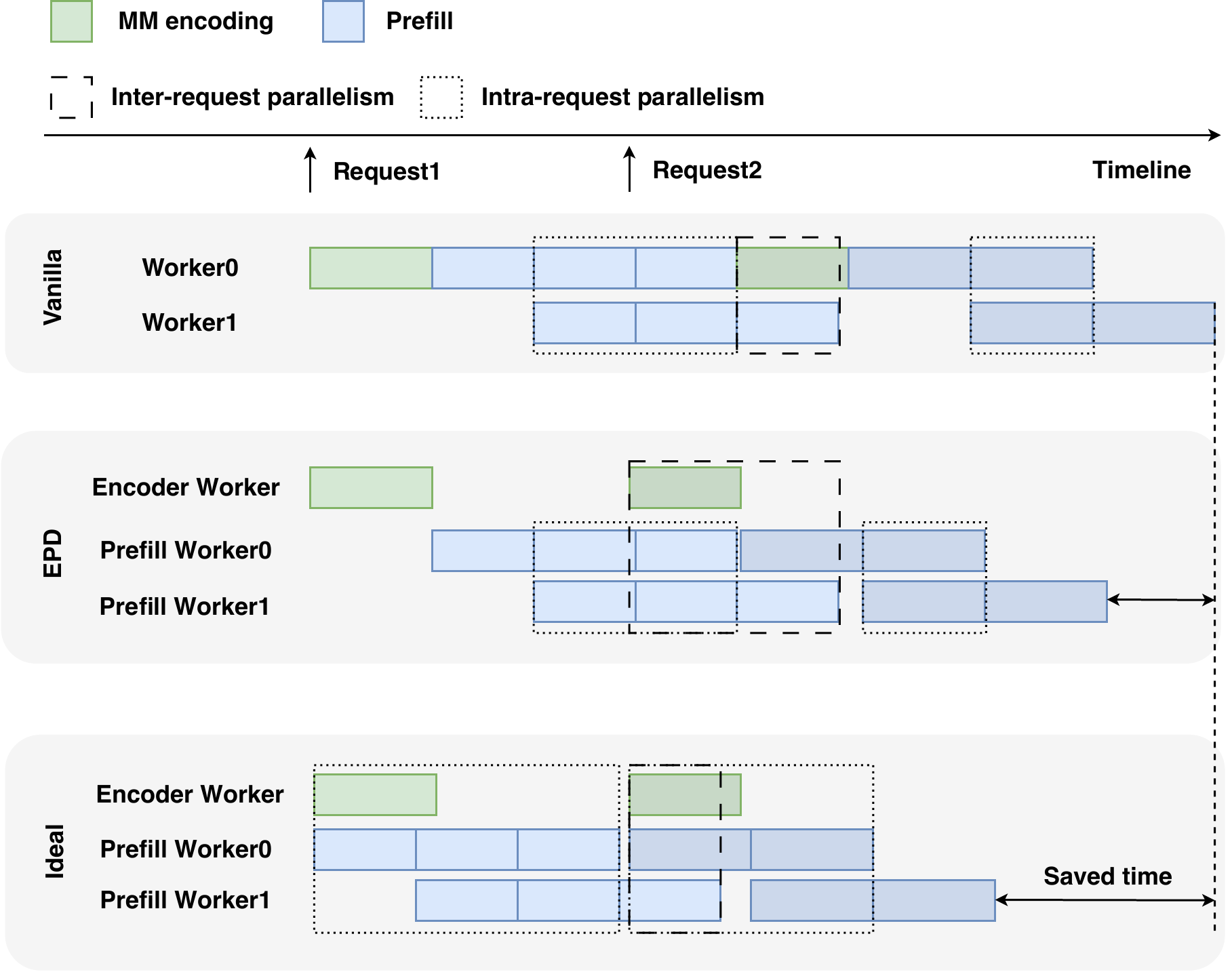}
    \caption{Different parallelism strategies, including vanilla pipeline parallelism (PP2), encoder-prefill (EP) \xwz{disaggregation}{disaggregated}, and ideal intra-request parallelism, result in noticeable differences in latency and resource utilization during LMM inference. The figure, showing two requests arriving at different times (lighter for Request 1, darker for Request 2), demonstrates the importance of efficient intra-request scheduling for improving performance.}
    \label{fig:irp}
\end{figure}

CPP leverages intra-request parallelism in LLM inference.
LMM introduces extra multimodal data encoding operations which can be naturally integrated into CPP.
When the preceding embedding is ready, we can start the prefill computation of ready embedding at once instead of waiting all multimodal embeddings in a request \xwz{to be}{have} finished.
In that case, multimodal encoding can be overlapped with prefill computation which further mitigates the time-consuming encoding operations.
However, there exists data dependency between encoding and CPP.
The image tokens must be encoded before the prefill operations.
Therefore, we should carefully interleave encoding and prefill operations.

Figure \ref{fig:irp} shows intra- and inter-request parallelism in LMM inference procedure. 
For the vanilla serving systems, the first worker is responsible for both the encoding and prefill operations.
The encoding \xwz{}{operation}and prefill \xwz{}{operation}interfere with each other.
For the encoder-disaggregated serving systems, the encoder worker is responsible for the multimodal encoding computation, while the remaining prefill workers execute the language model in pipeline fashion.
Inter-request parallelism occurs when encoding for \textit{Request2} overlaps with prefill computation for \textit{Request1}. 
Intra-request parallelism arises when encoding or prefill operations execute concurrently with other prefill computations within the same request.

\begin{figure*}[t]
    \centering
    \includegraphics[width=\linewidth]{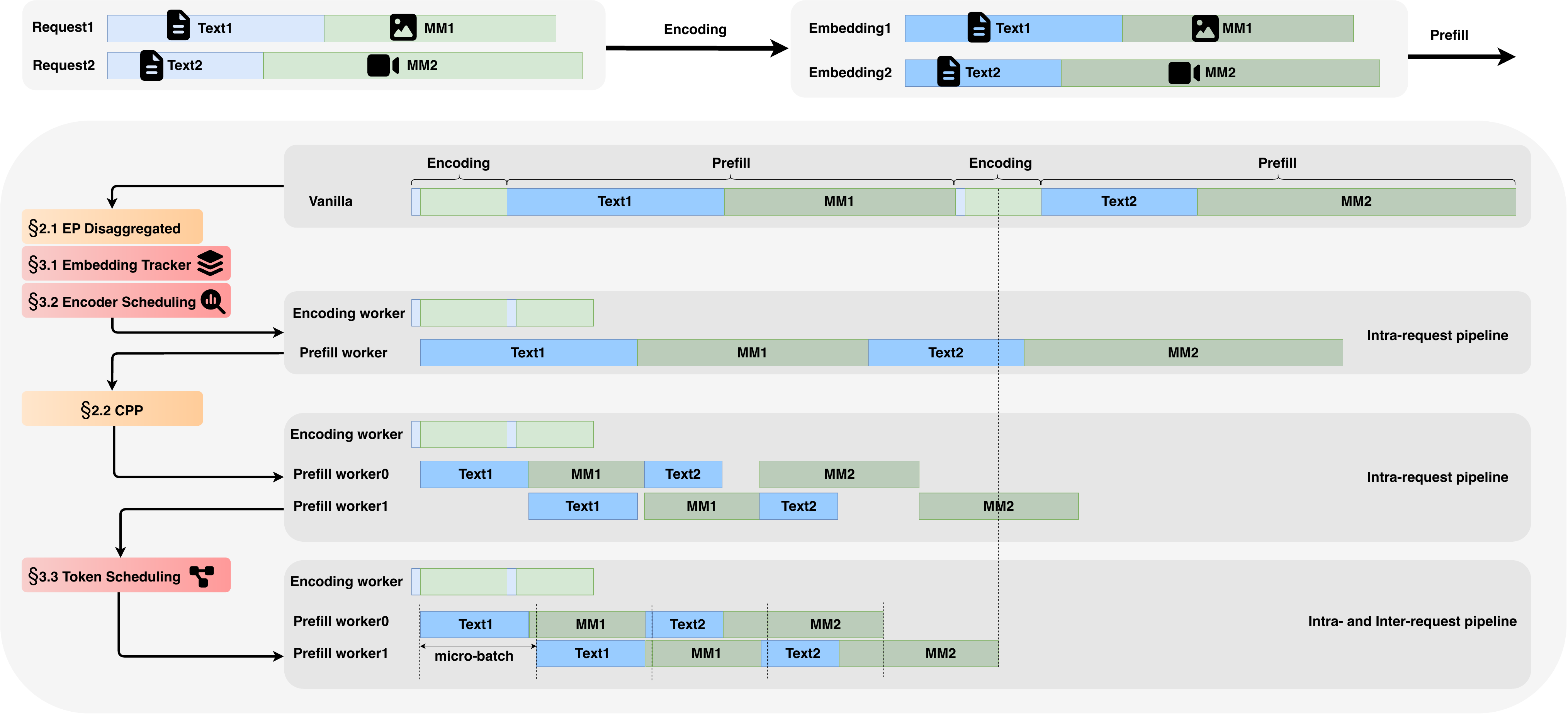}
    \caption{Overall effect of each component in \NAME. The EP disaggregated paradigm decouples encoding and prefill computations, enabling fine-grained overlap between these stages. By leveraging the embedding tracker and encoder scheduling, the system orchestrates concurrent execution of encoding and prefill operations. Furthermore, CPP minimizes the latency of individual requests through chunked pipelined execution. The introduction of schedulable tokens establishes both intra- and inter-request pipelines, maximizing utilization and overall execution efficiency.}
    \label{fig:overview}
\end{figure*}

To fully realize intra- and inter-request parallelism, several key challenges must be addressed:
(1) Determining prefill eligibility for embeddings: Prefill operations can only commence when the corresponding text or multimodal embeddings are ready, making dependency management critical;
(2) Managing embedding storage: Given the limited GPU memory, an efficient eviction strategy is \xwz{necessitated}{necessary} to remove unused embeddings while preserving computational efficiency;
(3) Encoding multimodal data in fine-grained granularity: The smaller the granularity of encoding, the greater the opportunity for overlapping computation; however, the computational efficiency of encoding decreases; 
(4) Scheduling prefill computation across multiple requests: This involves not only optimizing intra-request execution but also coordinating inter-request pipelines for improved throughput.

\section{DESIGN}

To better orchestrate intra- and inter-request pipeline\xwz{s}{}, we design \NAME, an efficient LMM serving system \xwz{to reduce}{for reducing} latency of rich multimodal requests. 
Built on the EP disaggregated architecture, \NAME~carefully organizes the execution of multimodal encoding and prefill computation by maintaining a per-request embedding tracker. 
The tracker indicates the ready embeddings for prefill execution and is in charge of releasing them once the corresponding prefill operation is completed.
To enable fine-grained overlapping, we schedule encoding computation in chunked granularity.
For \xwz{cooperating}{cooperated} multiple request scheduling, we propose a token scheduling method to manage the execution progress of different requests. 
Figure \ref{fig:overview} illustrates the effect of each \NAME~module on the latency and scheduling of LMM inference.

\subsection{Embedding Tracker for Intra-request Pipeline}

\begin{figure}[ht]
    \centering
    \includegraphics[width=\linewidth]{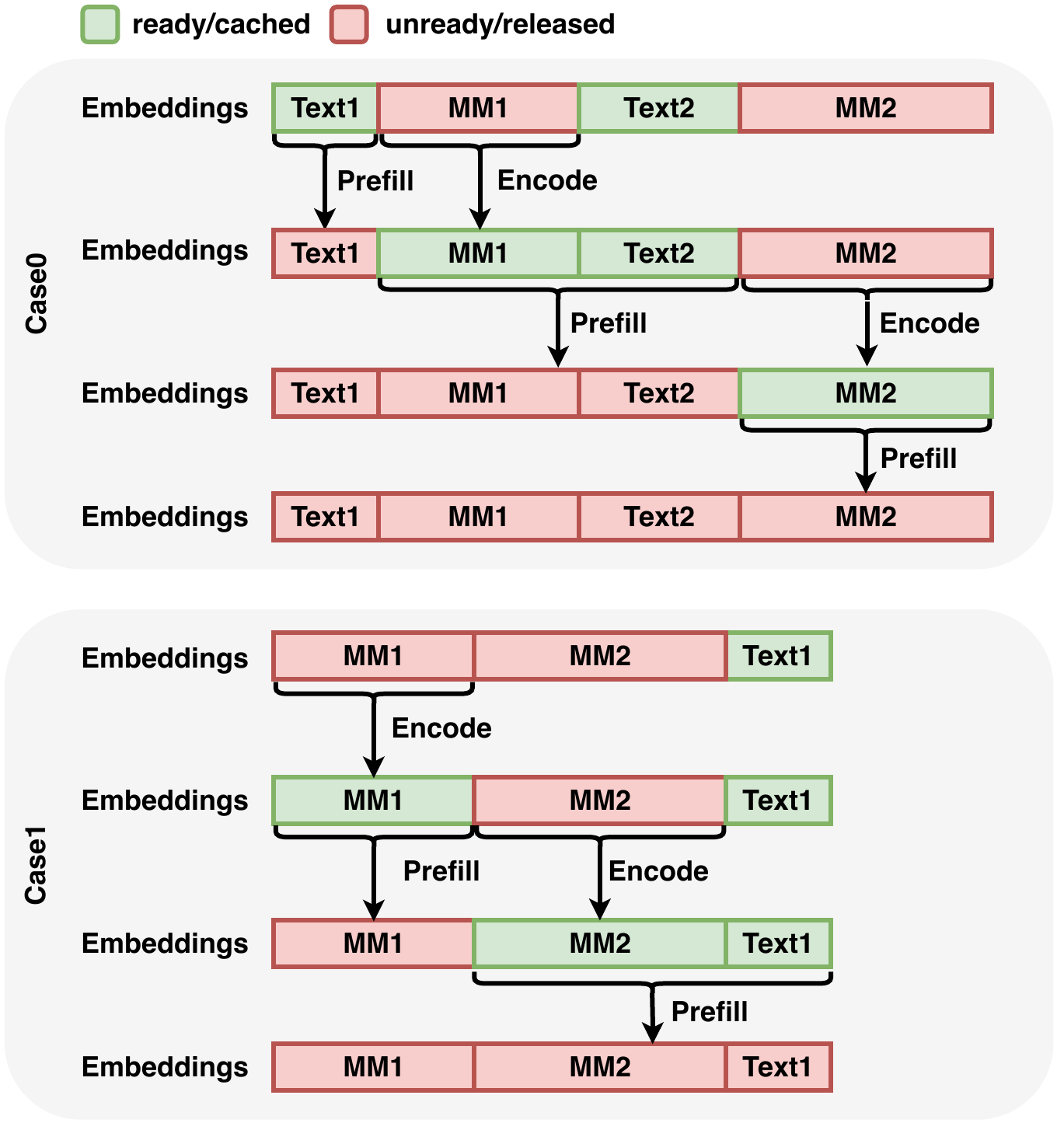}
    \caption{A series of encoding and prefill operations in the embedding tracker. The request contains both multimodal (MM) data and text data.}
    \label{fig:embedding-tracker}
\end{figure}

To coordinate the execution of multimodal encoding and prefill operations, \NAME~employs a per-request embedding tracker.
This tracker maintains the embeddings generated from multimodal data and manages their readiness for prefill computation.
When a new request is created, the embedding tracker initializes the \xwz{its}{request’s} metadata, including the embedding dimensions and readiness tags.
The embedding dimensions record the \xwz{token counts of}{number of tokens (} both text and multimodal\xwz{,}{)} and the hidden size of each token’s embedding.
Each readiness tag is set to \xwz{true}{True} for text tokens \xwz{with ready embeddings}{(since their embeddings are readily available)} and \xwz{false}{False} for multimodal tokens whose embeddings require further computation.
As multimodal embeddings are generated, the tracker updates their readiness tags and stores the new embeddings in their corresponding positions within the request.
Once embeddings have been passed to the LLM for prefill, the tracker immediately releases them to avoid memory leaks.
This mechanism ensures correct execution order and triggers prefill computation as soon as embeddings become available.

Figure \ref{fig:embedding-tracker} illustrates the workflow of the embedding tracker. 
When a request is created, \NAME~fetches all text embeddings upfront, whose cost is negligible, and reserves positions for multimodal embeddings.
In Case0, \NAME~first schedules the prefill for Text1 while concurrently performing the encoding of MM1.
Once MM1 is encoded, it triggers the prefill for MM1 and Text2, while simultaneously starting the encoding of MM2.
\xwz{When}{After} MM2 embeddings are ready, \NAME~executes the prefill for MM2.
After each prefill step, the corresponding embeddings are released to free GPU memory.
Through this fine-grained interleaving of encoding and chunked prefill, \NAME~maximizes GPU utilization and reduces latency.

\xwz{Particularly}{Importantly}, \NAME~does not require the input to follow a specific pattern, e.g., starting with text or alternating between multimodal and text segments.
As long as the input contains multiple multimodal elements, \NAME~can overlap encoding and prefill computations.
For instance, in Case1, where the input has two consecutive multimodal items (\textit{MM1} and \textit{MM2}), \NAME~can prefill \textit{MM1} while simultaneously encoding \textit{MM2}.

\subsection{Encoder Scheduling for Intra-request Pipeline}

\begin{algorithm}[ht]
\caption{Encoder Scheduling in LMM Inference}
\label{alg:encoder-scheduling}
\begin{algorithmic}[1]
\Require Waiting queue $Q$, Embedding batch size $C$
\While{True}
    \If{$Q$ is not empty}
        \State $req \gets$ Dequeue($Q$) \Comment{First come, first served}
        \State $buffer \gets \emptyset$
        \For{each element $e$ in $req$}
            \If{$e$ is multimodal data}
                \State Append $e$ to $buffer$
                \If{$|buffer| \geq C$}
                    \State Encode($buffer$)
                    \State $buffer \gets \emptyset$ \Comment{Reset $buffer$}
                \EndIf
            \EndIf
        \EndFor
        \If{$|buffer| > 0$}
            \State Encode($buffer$) \Comment{Remaining MM data}
        \EndIf
    \Else
        \State Wait for new request
    \EndIf
\EndWhile
\end{algorithmic}
\end{algorithm}

Current LMM serving systems usually batch all multimodal inputs within a request and process them together. While this strategy simplifies execution, it strictly enforces a dependency \xwz{that}{:} the prefill phase cannot start until all multimodal data \xwz{has}{have} been fully encoded. Such rigid ordering not only increases end-to-end latency but also prevents effective overlap between encoding and prefill computation.

To \xwz{tackle}{break} this bottleneck, encoding should ideally proceed in a streaming manner, where embeddings are generated and forwarded to the LLM for prefill as soon as they become available. This enables fine-grained pipeline parallelism between the encoder and the LLM worker. However, in practice, a single request often contains dozens of multimodal items \xwz{such as}{(e.g.,} multiple images, audio segments, or video frames. Encoding them strictly one by one leads to highly inefficient execution, as small batch sizes severely underutilize the GPU and make the encoder computation memory-bound.

To balance latency and hardware efficiency, \NAME~ado-pts a batching strategy for embeddings (Algorithm \ref{alg:encoder-scheduling}). Specifically, multimodal items are organized into batches containing at least $C$ multimodal tokens, and each batch is encoded together.
This ensures that the encoder achieves sufficient parallelism without waiting for the entire request to be ready. 
Since multimodal encoding cannot be divided at the token level, \NAME~treats each multimodal item as an indivisible execution unit and aggregates them into batches, enabling overlap with prefill computation while maintaining high encoding efficiency.

\subsection{Token Scheduling for Inter-request Pipeline}
\label{sec:inter-pipe}

\begin{algorithm}[ht]
\caption{CPP Scheduling with Schedulable Tokens}
\label{alg:intra-inter-request-scheduling}
\begin{algorithmic}[1]
\Require Waiting queue $Q$, Token budget $B$
\Ensure A batch of requests for execution
\State $S \gets \emptyset$ \Comment{Initialize scheduling queue}
\State $U \gets \emptyset$ \Comment{Initialize incomplete request queue}
\While{$Q \neq \emptyset$ \textbf{and} $B > 0$}
    \State $r \gets \text{Dequeue}(Q)$
    \State $t \gets \text{SchedulableTokens}(r)$
    \State $p \gets \text{PromptLength}(r)$
    \If{$t \le B$}
        \State Add $r$ to $S$
        \State $B \gets B - t$ \Comment{Update token budget}
    \Else
        \State Add $r$ to $S$
        \State $B \gets 0$ \Comment{Update token budget}
    \EndIf
    \If{$t < p$}
        \State Add $r$ to $U$ \Comment{Mark as incomplete}
    \EndIf
\EndWhile
\If{$S \neq \emptyset$}
    \State \text{BatchExecute}($S$) \Comment{LLM forward pass}
\EndIf
\If{$U \neq \emptyset$}
    \State \text{Prepend}($U$, Q) \Comment{Move incomplete requests to front}
\EndIf
\end{algorithmic}
\end{algorithm}

Apart from intra-request pipeline, modern LLM serving systems also explore inter-request pipeline, where multiple requests are batched together \xwz{with}{under} a shared token budget. However, when extending to LMM serving, this approach encounters new challenges.
Specifically, token scheduling cannot proceed until the corresponding multimodal embeddings have been generated, as prefill computation requires them.
This creates a tight data dependency between multimodal encoding and prefill scheduling, as prefill cannot be scheduled until the corresponding embeddings have been fully generated, which complicates the design of efficient scheduling policies.

To address this challenge, \NAME~introduces the concept of schedulable tokens (Algorithm \ref{alg:intra-inter-request-scheduling}). A token becomes schedulable once its multimodal embedding is ready and its preceding tokens have either completed prefill computation or themselves become schedulable. Based on this mechanism, \NAME~dynamically maintains a pool of schedulable tokens and uses a global token budget to batch them across different requests. During each scheduling iteration, tokens are dequeued, evaluated for eligibility, and \xwz{then}{} placed into the execution batch if the remaining token budget \xwz{permits}{allows}. Requests that cannot be fully scheduled are marked as incomplete and reinserted into the head of waiting queue with updated state, ensuring they will be revisited promptly in the next scheduling round.
By incrementally updating token states as soon as their embeddings are available, \NAME~enables efficient batching of prefills across heterogeneous multimodal requests. This mechanism not only supports intra-request pipeline, but also effectively constructs inter-request pipelines, thereby achieving both low latency and high throughput in LMM serving.

\begin{figure}[t]
    \centering
    \includegraphics[width=\linewidth]{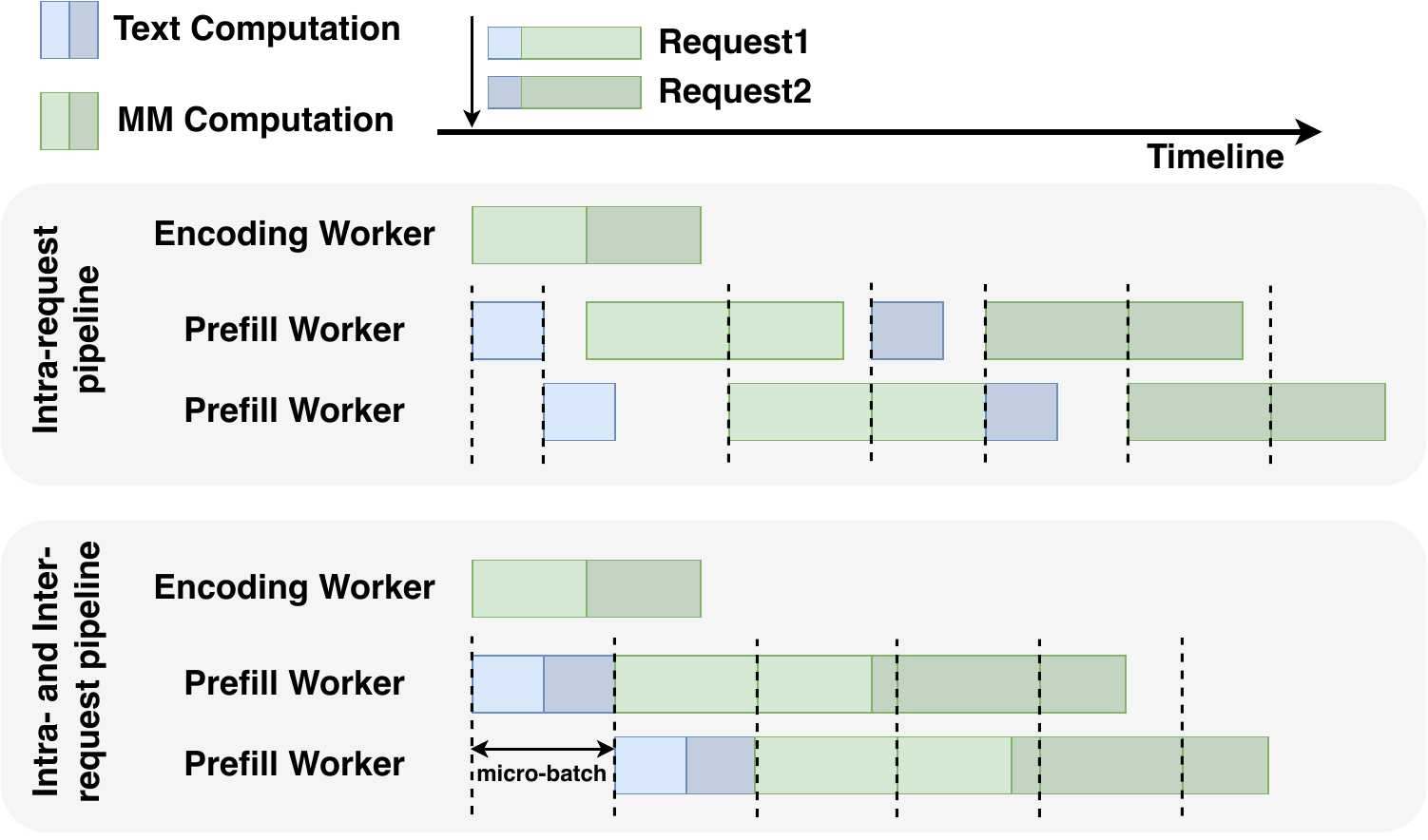}
    \caption{Comparison between the intra-request pipeline and the combined intra- and inter-request pipeline.}
    \label{fig:scheduling}
\end{figure}

\xwz{Putting together,}{} Figure \ref{fig:scheduling} \xwz{showcases}{illustrates} the benefits of adopting both intra-request and inter-request pipeline scheduling.
When relying solely on intra-request pipeline, the serving system struggles to fully utilize the available token budget within each micro-batch, which leads to significant pipeline bubbles and underutilization of computational resources.
In contrast, by combining intra-request and inter-request pipeline, the system can aggregate tokens from multiple requests to completely fill the micro-batch.
This strategy not only mitigates idle time in the pipeline, but also achieves a balanced trade-off between latency and throughput, enabling the system to simultaneously deliver low response time for individual requests while sustaining high overall throughput.

\subsection{Implementation}

\begin{figure}[ht]
    \centering
    \includegraphics[width=\linewidth]{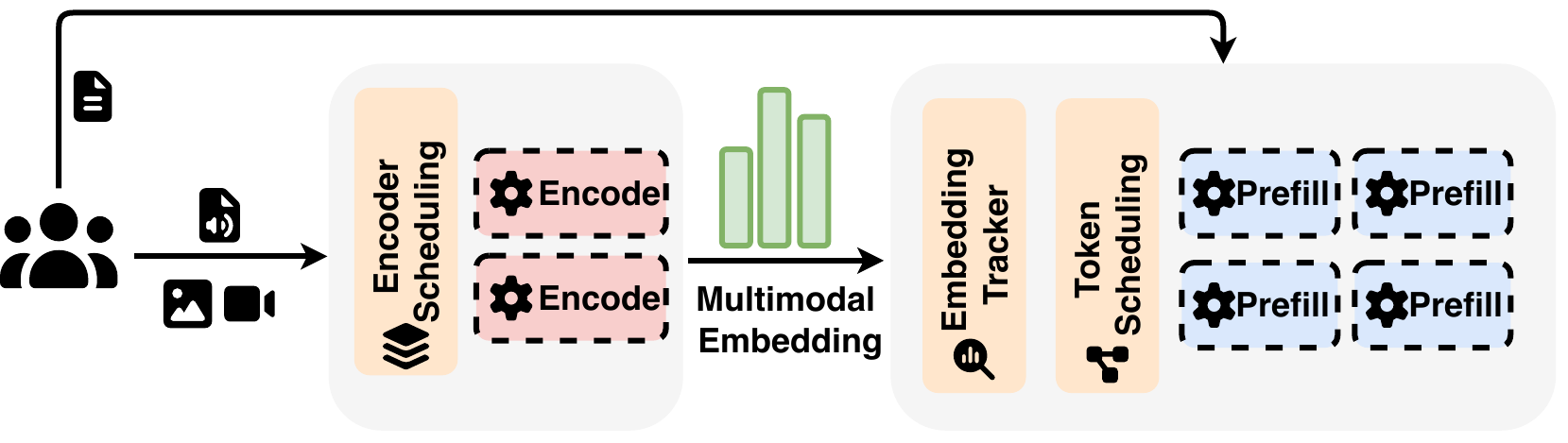}
    \caption{\xwz{Overall workflow}{Overview} of \NAME~prototype.}
    \label{fig:overview-workflow}
\end{figure}

We implement \NAME, with the component diagram being shown in Figure \ref{fig:overview-workflow}, on top of gLLM \cite{gLLM}, \xwz{which is}{} a lightweight and highly efficient LLM/LMM serving framework, \xwz{achieving}{which achieves} performance comparable to vLLM while maintaining a simpler and more flexible architecture.
The embedding tracker is based on a dictionary data structure, where each request ID is used as the key and the corresponding embedding cache is stored as the value. 
The driver worker is responsible for maintaining this tracker and orchestrating prefill scheduling. 
Specifically, the tracker is used to determine the number of schedulable tokens for each request and to prepare the model input embeddings. 
When the driver worker receives new embeddings generated by the model, it updates the corresponding cache entry in place so that scheduling decisions always rely on the latest embedding states.
In addition, we modify the gLLM scheduler to support token-level scheduling. 
Instead of scheduling entire requests, the scheduler operates on the number of available schedulable tokens.

\begin{figure*}[ht]
    \begin{subfigure}[b]{.48\linewidth}
        \centering
        \includegraphics[width=\linewidth]{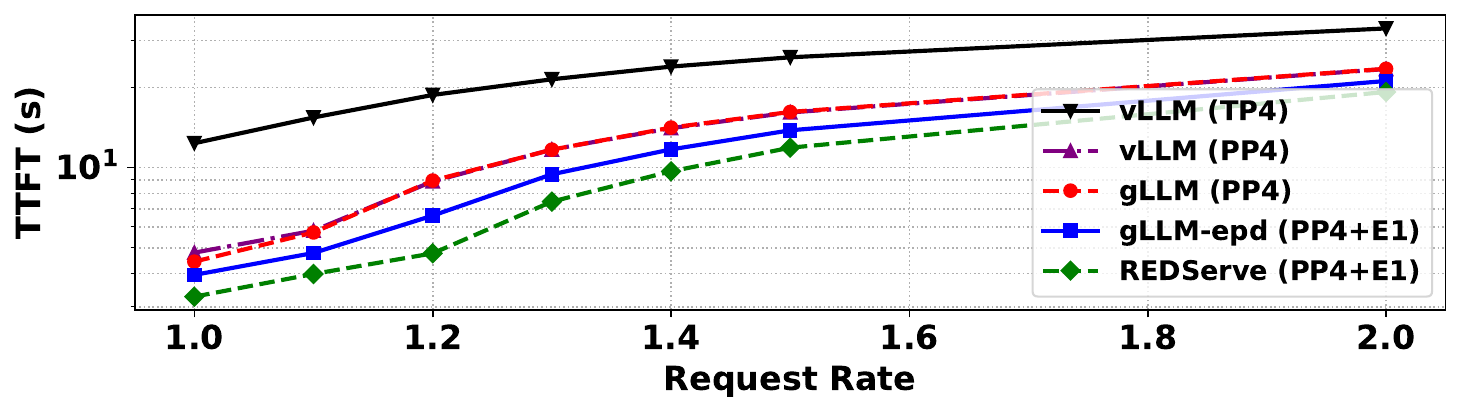}
        \caption{Qwen2.5-VL-72B, 1K resolution.}
    \end{subfigure}
    \begin{subfigure}[b]{.48\linewidth}
        \centering
        \includegraphics[width=\linewidth]{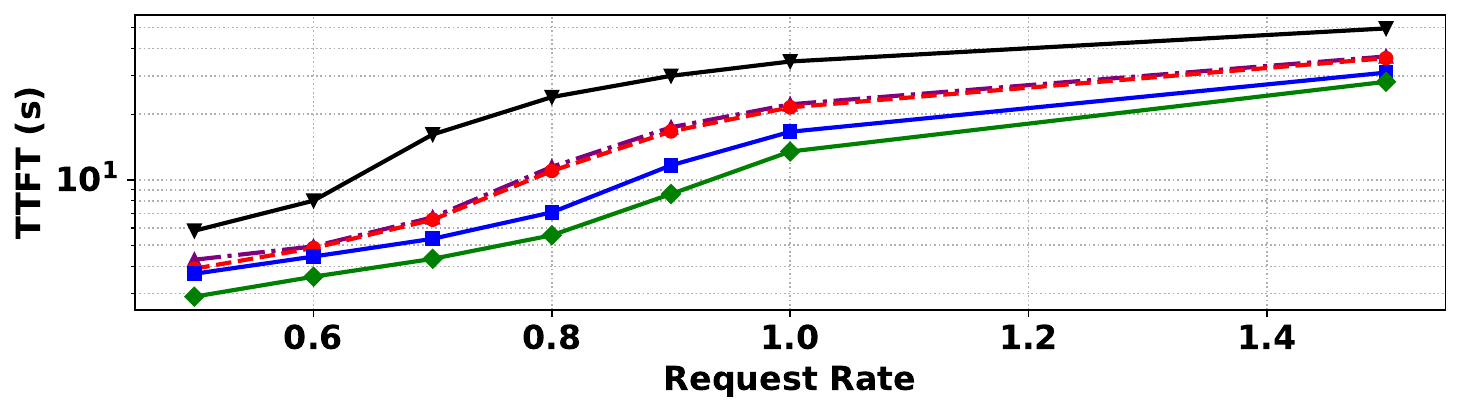}
        \caption{Qwen2.5-VL-72B, 2K resolution.}
    \end{subfigure}
    \caption{Latency comparison of vLLM, gLLM and \NAME~(logarithmic coordinate system).}
    \label{fig:res-TTFT}
\end{figure*}

\begin{figure*}[ht]
    \begin{subfigure}[b]{.48\linewidth}
        \centering
        \includegraphics[width=\linewidth]{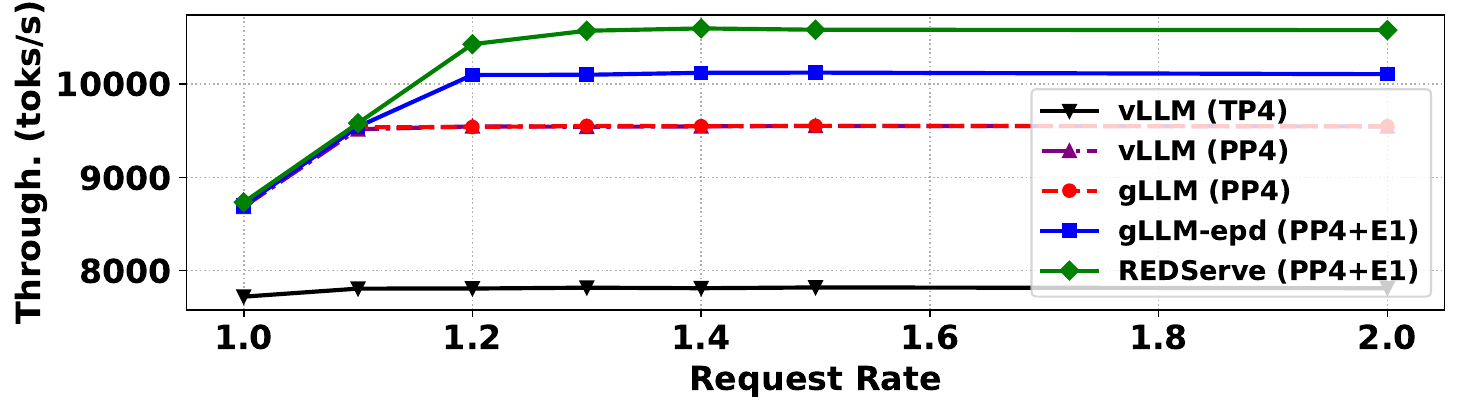}
        \caption{Qwen2.5-VL-72B, 1K resolution.}
    \end{subfigure}
    \begin{subfigure}[b]{.48\linewidth}
        \centering
        \includegraphics[width=\linewidth]{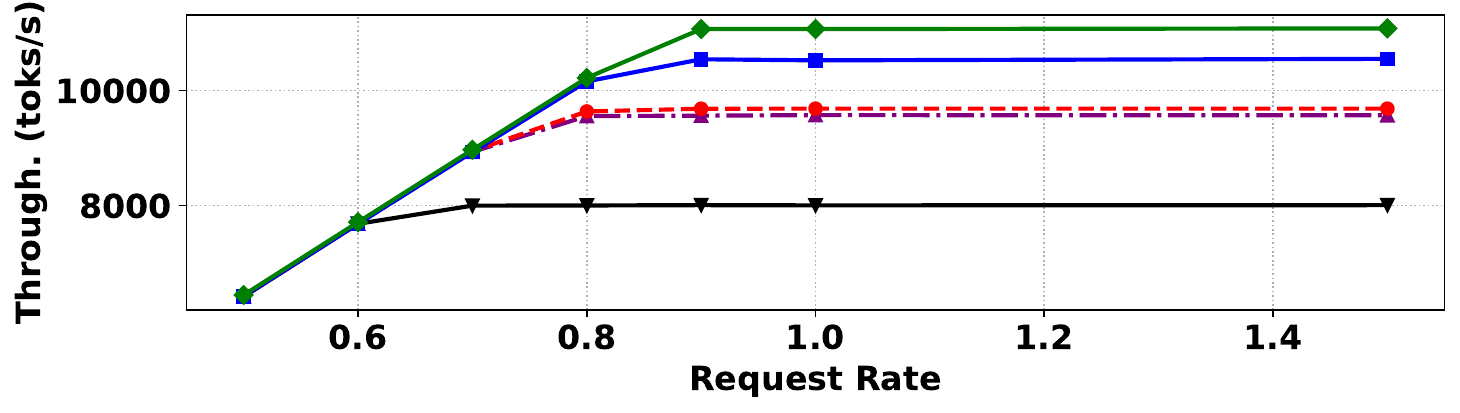}
        \caption{Qwen2.5-VL-72B, 2K resolution.}
    \end{subfigure}
    \caption{Throughput comparison of vLLM, gLLM and \NAME.}
    \label{fig:res-through}
\end{figure*}

\begin{figure*}[ht]
    \begin{subfigure}[b]{.48\linewidth}
        \centering
        \includegraphics[width=\linewidth]{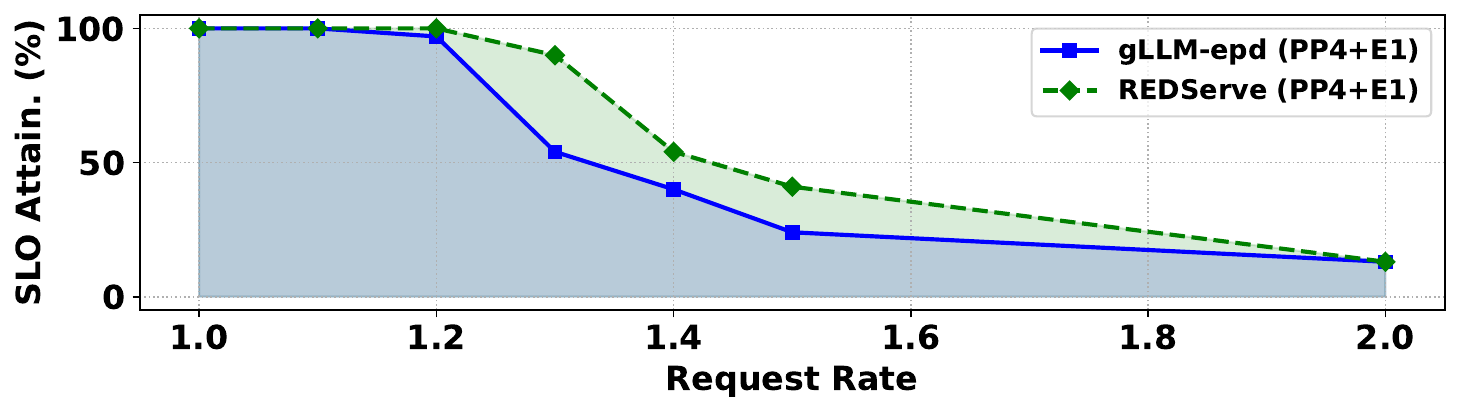}
        \caption{Loose SLO TTFT:10s, Qwen2.5-VL-72B, 1K resolution.}
    \end{subfigure}
    \begin{subfigure}[b]{.48\linewidth}
        \centering
        \includegraphics[width=\linewidth]{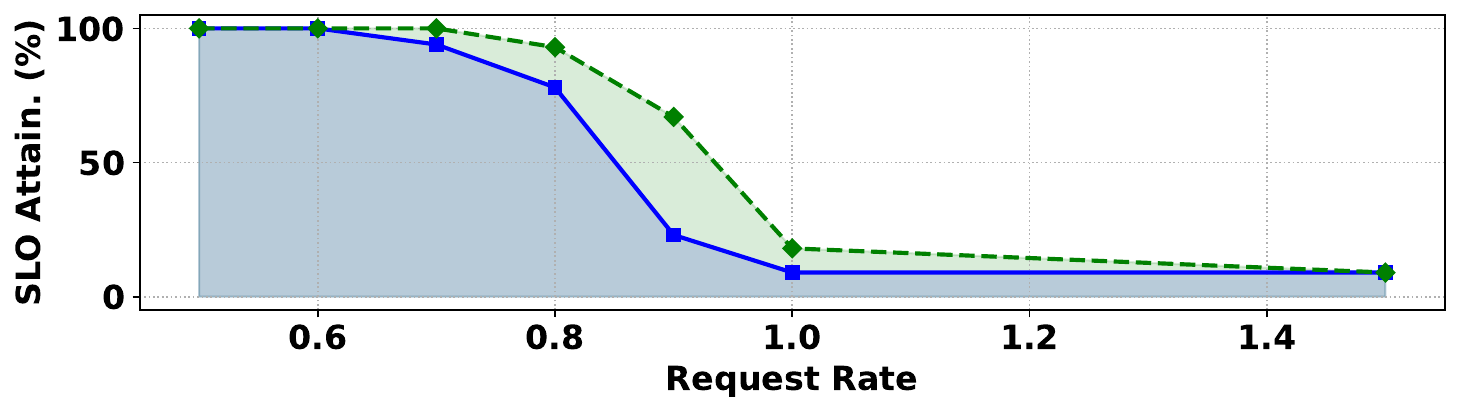}
        \caption{Loose SLO TTFT:10s, Qwen2.5-VL-72B, 2K resolution.}
    \end{subfigure}
    \caption{SLO Attainment comparison of gLLM and \NAME.}
    \label{fig:res-goodput}
\end{figure*}

\xwz{\NAME~is applicable to diverse scenarios. In particular}{
Within \NAME}, the intra-request pipeline optimization is orthogonal to existing parallelism strategies.
It can be seamlessly integrated with both pipeline parallelism and tensor parallelism, and is equally applicable to single-GPU model deployments, enabling deployment flexibility across different hardware scales.
In contrast, the inter-request pipeline is inherently coupled with pipeline parallelism; it acts as a hybrid mechanism that fuses \NAME’s intra-request pipeline optimization with pipeline parallel execution to further enhance system throughput and reduce end-to-end latency under multi-request workloads.

\section{EVALUATION}

\subsection{Experimental Setup}

Our experiments are performed under an EPD disaggregated configuration. 
As encoding and prefill operations predominantly influence the time-to-first-token (TTFT), our evaluation emphasizes first-token latency rather than inter-token latency. 
To accurately simulate the behaviour of prefill and encoding nodes, we fix the output length to one and collect TTFT or throughput as the primary performance metrics.

\subsubsection{Models and Environments}
We evaluate \NAME\\using the Qwen2.5-VL \cite{Qwen2_5_VL} series (7B, 32B and 72B variants), \xwz{considering}{chosen for} their strong multimodal capabilities. 
The main experiments are conducted on a system equipped with a 140-core Intel(R) Xeon(R) processor (1.37 TB host memory) and 8$\times$H100 GPUs (80 GB each) connected by NVLink.
To further verify REDServe's robustness, we also evaluate it on a system equipped with a 64-core AMD EPYC 7742 processor (256 GB host memory) and 4$\times$A100 GPUs (40GB each) connected by PCIE (\S \ref{sec:a100}).
The experiments are conducted using Python, version 3.12.11.

\subsubsection{Workloads}

\begin{figure}[t]
    \centering
    \includegraphics[width=\linewidth]{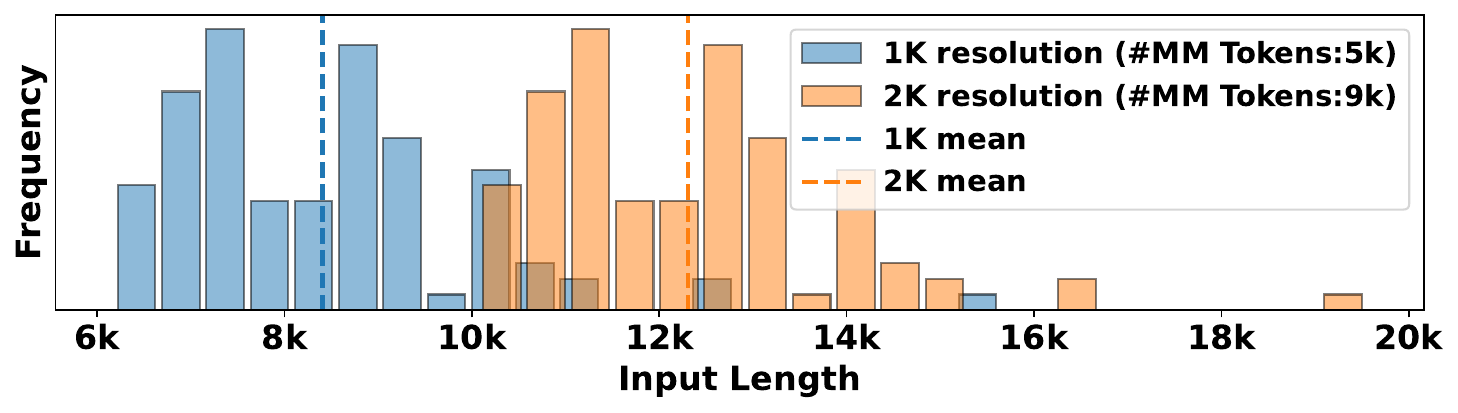}
    \caption{Distribution of input lengths for different resolutions in MMMU.
    For 1K and 2K resolutions, the number of multimodal tokens is 5k and 9k, respectively.}
    \label{fig:input_dis}
\end{figure}

We construct workloads using the dataset MMMU \cite{MMMU} and the open-source benchmark in SGLang \cite{SGLang}. 
MMMU is a large-scale multimodal benchmark spanning diverse domains such as science, engineering, and humanities, featuring text, images, charts, and diagrams that require expert-level reasoning. 
To emulate a cloud service environment, we generate request arrivals following a Poisson distribution with varying rates as in vLLM \cite{vllm}.
We vary the image resolution to emulate diverse multimodal workloads.
Figure \ref{fig:input_dis} shows the distribution of input lengths in MMMU.
For 1K and 2K resolutions, the average input length is 8k and 12k.

\subsubsection{Schemes} 
We benchmark \NAME~with following systems.
\begin{itemize}
    \item \textbf{vLLM \cite{vllm}.} We use vLLM (v0.10.1.1) as a baseline to gauage \NAME's performance. 
    As one of the most widely adopted inference engines, vLLM is renowned for its rich feature set and extensive model support.
    We have applied a critical bug fix (\#24387) to address issues in the multimodal encoder within the pipeline parallelism setup.
    \item \textbf{gLLM \cite{gLLM}.} We use gLLM (v0.0.4) both as a performance baseline and as the foundation for our implementation.
    gLLM is a lightweight inference system specifically designed for rapid development and experimental validation.
    \item \textbf{gLLM-epd\footnote{Since there is currently no mature implementation of EP disaggregation in existing open-source frameworks, we implemented a EP-disaggregated version based on the gLLM framework, and our experiments demonstrate that it achieves the expected performance.}.} EP disaggregated version developed based on gLLM.  
    \item \textbf{\NAME.} Proposed efficient LMM serving system by orchestrating intra- and inter-request pipeline. 
    The implementation is based on gLLM-epd.
    \item \textbf{\NAME-intra.} \NAME~without inter-request pipeline (\S \ref{sec:inter-pipe}).
\end{itemize}
All the schemes employ \CP's scheduling strategy \cite{Sarathi-serve} and the token budget remains consistent.
To eliminate the impact of KV cache reuse on performance, we disable cross-request KV cache reuse for all schemes.
The GPU memory utilization of each system is
set to the maximum without encountering out of memory error.
For pipeline parallelism, CPP is enabled by default.
For \NAME, we set the embedding batch size to 1024. 
In the Section \S \ref{sec:embedding_chunk_size}, we investigate how varying parameter settings affect performance.

\subsubsection{Metrics} \xwz{We consider the following evaluation metrics:}{}
\begin{itemize}
    \item \textbf{Time to First Token (TTFT).} Average time taken from when a user sends a prompt to the LMM until the first token of the response is generated.
    \item \textbf{Throughput.} Average input tokens processing throughput.
    \item \textbf{SLO Attainment \cite{DistServe}.} The SLO fulfilment rate under the given TTFT constraint.

\end{itemize}

\subsection{\xwz{Performance Improvement}{}}

Pipeline parallelism pairs well with \NAME. We begin by evaluating \NAME’s performance under this configuration, as illustrated in Figures \ref{fig:res-TTFT}, \ref{fig:res-through}, and \ref{fig:res-goodput}.

\subsubsection{Latency}

To evaluate the performance of \NAME, we compare it against vLLM and gLLM under different parallelism strategies and serving architectures, as illustrated in Figure \ref{fig:res-TTFT}.
vLLM (TP4), which adopts tensor parallelism, suffers from significantly higher latency (up to 3.77$\times$) compared to systems based on pipeline parallelism. 
This overhead mainly arises from frequent synchronous communication in tensor parallelism, which severely degrades overall system performance.
In contrast, pipeline parallelism, especially when combined with CPP, not only increases throughput but also reduces per-request latency. 
The results show that TTFT of gLLM is very close to that of vLLM (PP4), with an average performance fluctuation of only 1.6\%/3.8\%, demonstrating that gLLM can achieve performance comparable to vLLM (PP4).
By further adopting an EP disaggregated architecture, gLLM-epd achieves an additional 16\%/20\% reduction in TTFT compared to gLLM. 
The performance advantage of gLLM-epd over gLLM follows an initial increase followed by a decrease (from 5\%/10\% to 26\%/35\% and then down to 10\%/14\%). 
This is because, at low request rates, gLLM-epd cannot effectively reduce latency, while at higher request rates, latency deteriorates significantly due to heavy request backlogs, thereby diminishing its advantage.
Building upon this, \NAME~achieves another 18\%/19\% reduction in TTFT compared to gLLM-epd by fully leveraging intra-request parallelism between multimodal encoding and LLM forward passes. 
\xwz{Putting together}{As a result}, \NAME~is particularly effective under low request rates, where intra-request parallelism dominates. 
As the request rate increases, \xwz{\NAME's}{its} performance gradually converges with that of gLLM-epd.

\subsubsection{Throughput}

We further evaluate the input token processing throughput of vLLM, gLLM, and \NAME~as shown in Figure \ref{fig:res-through}. 
As the request rate increases, throughput initially rises and then stabilizes, with the plateau representing the maximum capacity the system can sustain. 
The tensor-parallel system represented by vLLM (TP4) exhibits significantly lower (26\%/28\%) throughput than pipeline-parallel systems, a trend consistent with the observed latency results. 
This again confirms that CPP empowers pipeline parallelism to surpass tensor parallelism on performance. 
Meanwhile, gLLM and vLLM (PP4) demonstrate nearly identical throughput performance (gap less than 1.2\%). 
With the integration of EPD, gLLM-epd achieves an additional throughput improvement of 6\%/8.5\% over gLLM. 
By exploiting both intra- and inter-request parallelism, \NAME~further extends the throughput limit, reaching about 10600/11100 tokens/s.

\subsubsection{SLO Attainment}

We also evaluate the fulfillment rate of service metrics as shown in Figure \ref{fig:res-goodput}.
We can find that as the request rate increases SLO attainment gradually drops from 100\% to less than 20\%.
This is because as the request rate continues to increase, the queuing time of requests also grows, and the inference system gradually fails to meet the service requirements of \xwz{partial}{some} requests.
\NAME~maintains a higher SLO attainment (average is 71\%/70\%) compared to gLLM-epd (average is 61\%/59\%) due to overlapped computation between encoding and prefill operations.
The larger the covered area under the curve in the line chart, the higher the SLO satisfaction rate of the system across different request rates.
\NAME~achieves an 23\%/23\% larger coverage area than gLLM-epd.
This further demonstrates that \NAME~has stronger scheduling performance compared to gLLM-epd.

\subsection{\xwz{Performance Dissecting}{Sensitivity and Ablation Study}}

\subsubsection{Embedding Batch Size}
\label{sec:embedding_chunk_size}

\begin{figure}
    \begin{subfigure}[b]{\linewidth}
        \centering
        \includegraphics[width=\linewidth]{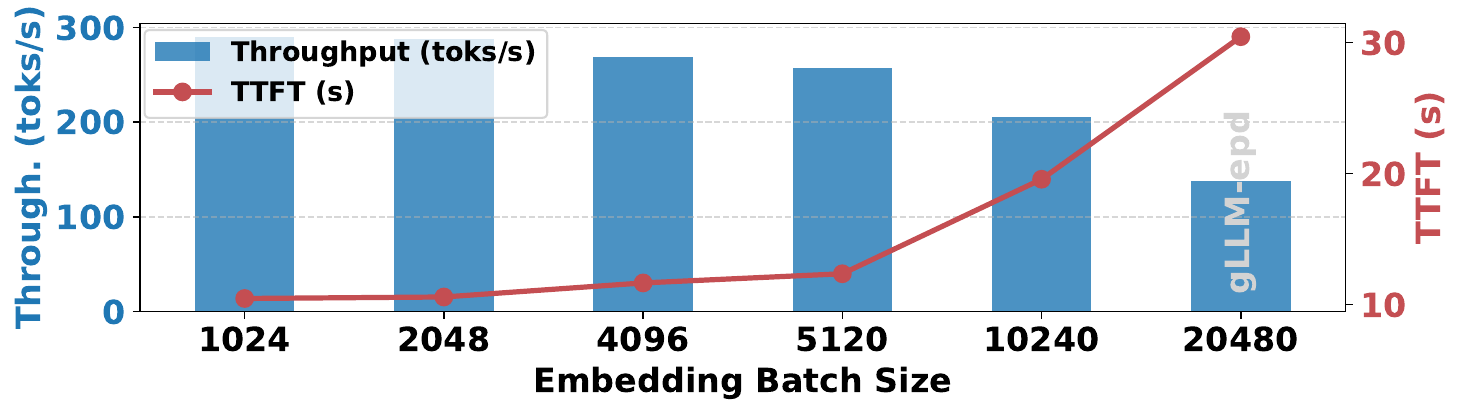}
        \caption{High-quality multimodal data (1024 toks/MM item)}
    \end{subfigure}
    \begin{subfigure}{\linewidth}
        \centering
        \includegraphics[width=\linewidth]{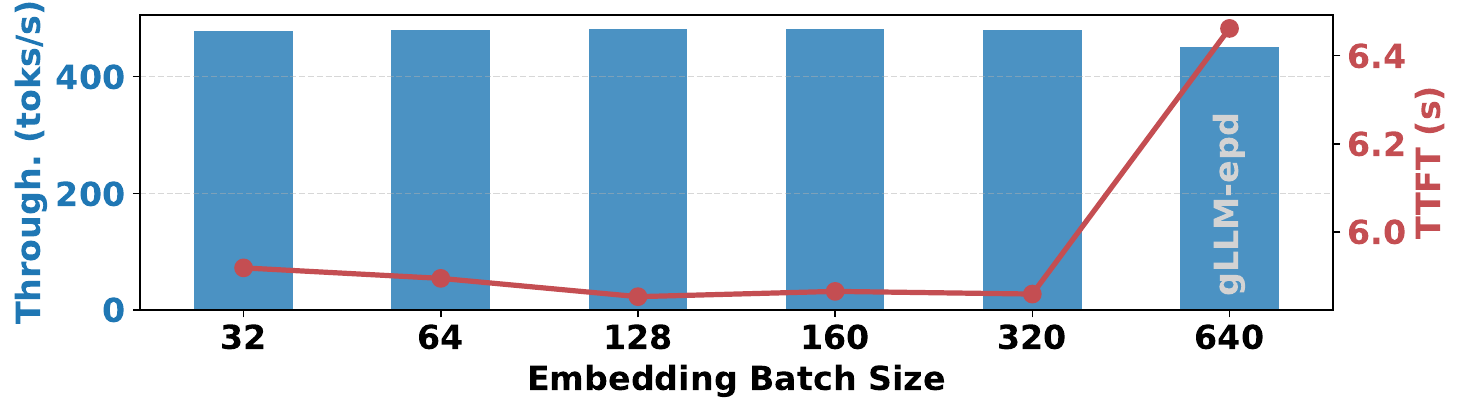}
        \caption{Low-quality multimodal data (32 toks/MM item)}
    \end{subfigure}
    \caption{Impact of varying embedding batch size. Two requests with about 2k text tokens and 20 MM items are sent to \NAME~(PP4+E1). The data on the far right represents performing the prefill operation only after completing the encoding of all multimodal data, which is equivalent to gLLM-epd.}
    \label{fig:embedding_chunk_size}
\end{figure}

In this section, we analyse the impact of different embedding batch sizes under both high-quality and low-quality multimodal data. The results are presented in Figure \ref{fig:embedding_chunk_size}.

For high-quality multimodal data, increasing the embedding batch size leads to a gradual rise in TTFT (by up to 2.91$\times$) and a steady decline in throughput (by as much as 53\%). 
This demonstrates that \NAME~can efficiently overlap encoding computation with prefill, and that finer-grained scheduling provides greater opportunities and longer durations for such overlap, ultimately improving system efficiency. 
Notably, even a single multimodal element is sufficient to fully utilize encoding computation capacity under this setting. 
Therefore, for high-quality multimodal data, a smaller embedding batch size is generally more advantageous.

For low-quality multimodal data, TTFT follows a different trend: it first decreases and then increases as the embedding batch size grows. 
This behaviour reflects the inherent trade-off between encoding efficiency and overlapped execution. 
At smaller batch sizes, execution time is dominated by encoding inefficiency. 
However, when the batch size grows too large, the opportunities for overlap diminish, and TTFT rises again. 
Hence, in practical deployment scenarios, the choice of embedding batch size should consider the balance between encoding efficiency and overlap benefits.

\subsubsection{Inter-request Pipeline}

\begin{figure}[t]
    \centering
    \includegraphics[width=\linewidth]{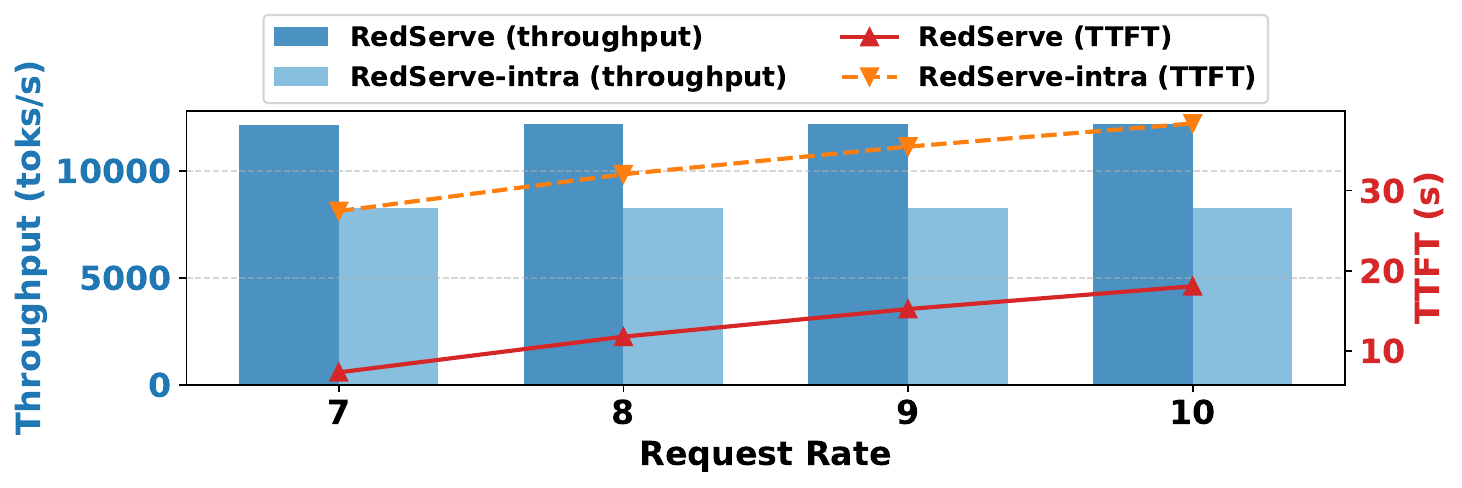}
    \caption{Ablation study between \NAME~and \NAME-intra on TTFT and throughput comparison.}
    \label{fig:inter-ablation}
\end{figure}

To evaluate the impact of the inter-request pipeline, we compare \NAME~with \NAME-intra (incorporating only intra-request pipeline), as shown in Figure~\ref{fig:inter-ablation}.
As the request rate increases, the throughput of both systems remains roughly constant, while the latency gradually rises. 
This is because the incoming request rate has already exceeded the maximum processing capacity of the systems.
\NAME-intra delivers 32\% lower throughput and 172\% higher latency than \NAME.
The absence of the inter-request pipeline significantly reduces the system’s processing speed, with the degradation in TTFT primarily attributed to longer waiting times.

\begin{figure*}[ht]
    \begin{subfigure}[b]{.48\linewidth}
        \centering
        \includegraphics[width=\linewidth]{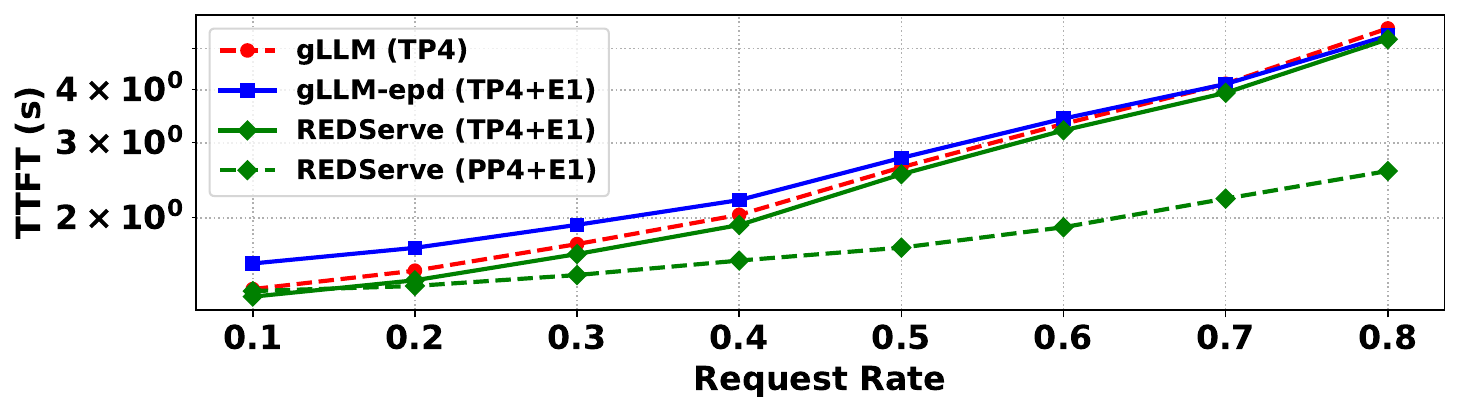}
        \caption{Qwen2.5-VL-72B, 1K resolution.}
    \end{subfigure}
    \begin{subfigure}[b]{.48\linewidth}
        \centering
        \includegraphics[width=\linewidth]{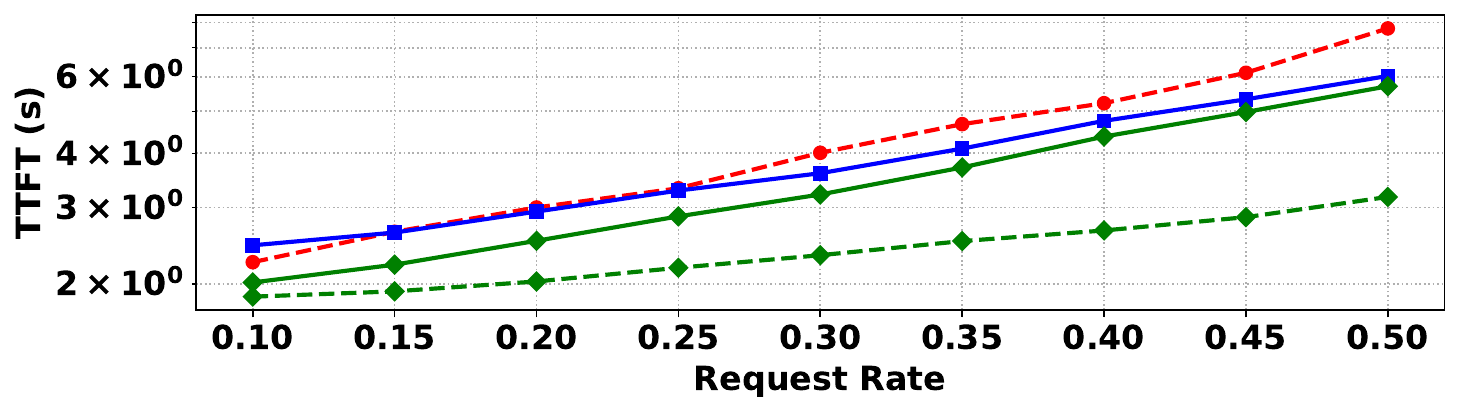}
        \caption{Qwen2.5-VL-72B, 2K resolution.}
    \end{subfigure}
    \caption{Latency comparison of gLLM and \NAME~(logarithmic coordinate system).}
    \label{fig:res-TTFT-TP}
\end{figure*}

\begin{figure*}[ht]
    \begin{subfigure}[b]{.48\linewidth}
        \centering
        \includegraphics[width=\linewidth]{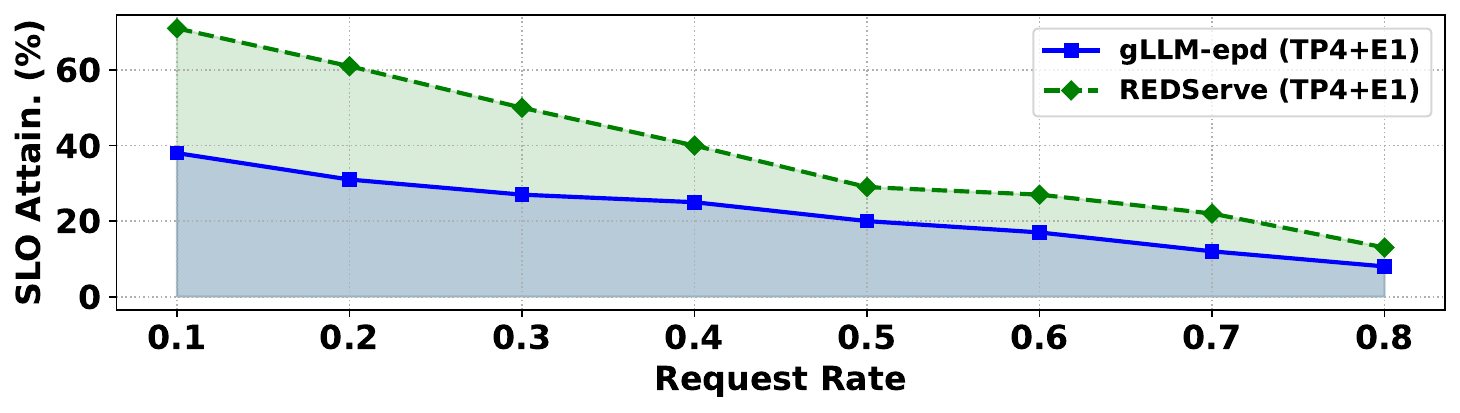}
        \caption{Strict SLO TTFT:1.4s, Qwen2.5-VL-72B, 1K resolution.}
    \end{subfigure}
    \begin{subfigure}[b]{.48\linewidth}
        \centering
        \includegraphics[width=\linewidth]{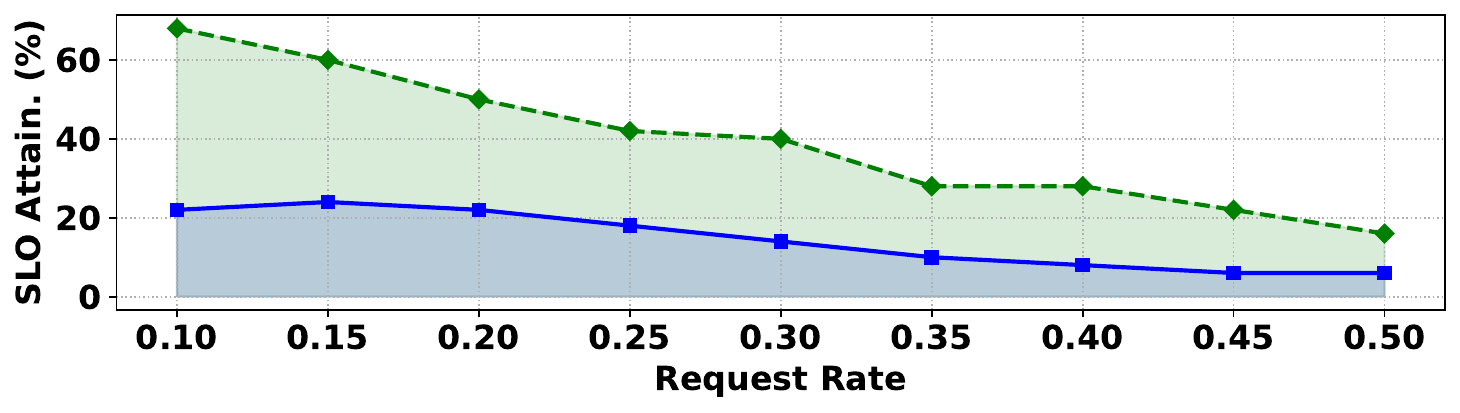}
        \caption{Strict SLO TTFT:2s, Qwen2.5-VL-72B, 2K resolution.}
    \end{subfigure}
    \caption{SLO Attainment comparison of gLLM and \NAME.}
    \label{fig:res-goodput-TP}
\end{figure*}

\subsection{Extensive \xwz{Studies}{Study}}

\subsubsection{\NAME~with Tensor Parallelism}
Tensor parallelism can also be integrated with \NAME. Accordingly, we further evaluate the serving performance of \NAME~when combined with tensor parallelism, as illustrated in Figure \ref{fig:res-TTFT-TP} and Figure \ref{fig:res-goodput-TP}.

\textbf{Latency.}
To evaluate our design under tensor parallelism, we compare gLLM and \NAME~across different architectures and parallelism strategies, as shown in Figure~\ref{fig:res-TTFT-TP}. 
We observe that the EPD architecture is not always beneficial: at low request rates, the additional embedding transmission overhead can actually increase latency. 
However, as the request rate grows, EPD becomes more effective, leveraging inter-request parallelism. \NAME~(TP4+E1) consistently outperforms both gLLM and gLLM-epd, with the performance gap over gLLM-epd narrowing at high request rates. 
This trend arises because \NAME~is particularly effective at reducing latency under low request rates. 
Notably, when combined with pipeline parallelism, \NAME~(PP4+E1) shows a clear latency advantage even as the request rate increases.

\textbf{Throughput.}
The evaluated throughput shows that when the request rate is relatively low, all schemes achieve almost identical performance, with throughput increasing approximately linearly with the request rate. 
This behaviour occurs because throughput is primarily constrained by the request arrival rate under such conditions. 
As the request rate rises, the performance of different schemes gradually diverges, reflecting differences in their efficiency and scalability under higher load.

\textbf{SLO Attainment.}
We evaluate gLLM-epd and \NAME~under strict SLOs, with the results presented in Figure \ref{fig:res-goodput-TP}. 
\NAME~consistently outperforms gLLM-epd, achieving over 75\%/169\% coverage. 
However, its advantage progressively declines with increasing request rates, mirroring the trend observed in latency.

\subsubsection{\xwz{\NAME~with Varied Settings.}{}} To validate its broad applicability, we evaluate \NAME~under several additional settings.

\begin{figure}[t]
    \centering
    \includegraphics[width=\linewidth]{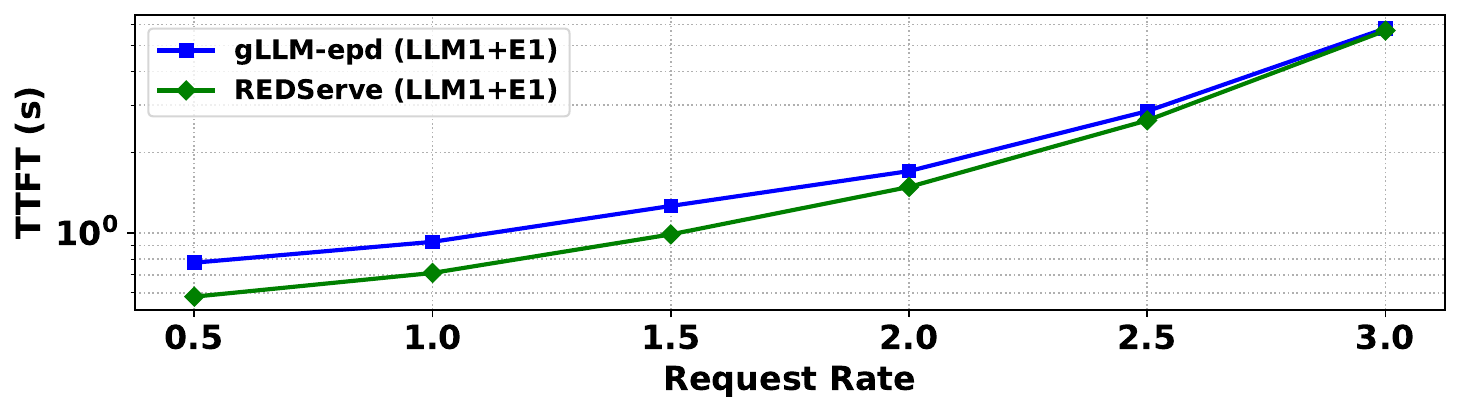}
    \caption{TTFT comparison of gLLM-epd and \NAME~when serving with one GPU for LLM and one GPU for multimodal encoding.}
    \label{fig:single}
\end{figure}

\textbf{Single-GPU deployment for LLM}. As shown in Figure \ref{fig:single}, \NAME~can also enhance the performance (up to 26\% TTFT reduction) for single-GPU deployment for LLM.
When the request rate is relatively low, the advantage of \NAME~becomes more pronounced, which is consistent with previous findings.

\label{sec:a100}

\begin{figure}[t]
    \centering
    \includegraphics[width=\linewidth]{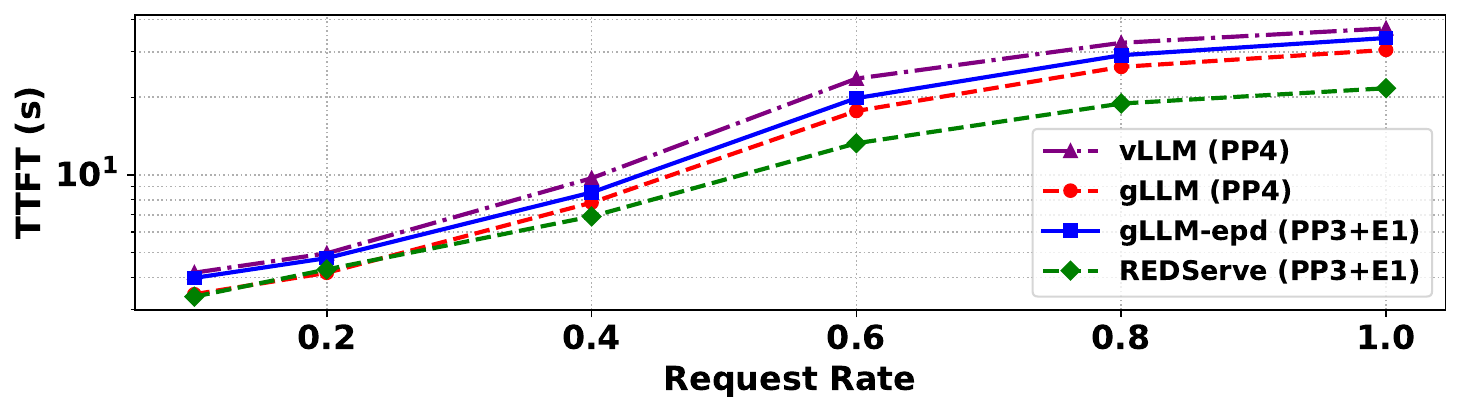}
    \caption{TTFT comparison of vLLM, gLLM and \NAME~on 4$\times$A100 GPUs.}
    \label{fig:a100}
\end{figure}

\textbf{A100 GPU Evaluations}. Under A100 GPUs, gLLM-epd fails to exhibit obvious performance advantage over gLLM.
However, \NAME~further fully leverage the parallelism potential and achieves the optimum performance. 

\subsubsection{Functional Study}

\begin{table}[t]
\centering
\caption{MMMU benchmark results of different inference approaches.}
\label{tab:mmmu_results}
\begin{tabular}{lcccc}
\toprule
\textbf{Framework} & vLLM & gLLM & gLLM-epd & \NAME \\
\midrule
\textbf{MMMU Score} & 62.7 & 62.6 & 62.4 & 62.6 \\
\bottomrule
\end{tabular}
\end{table}

To evaluate the functional usability of inference frameworks, we assess the inference performance of each system on the MMMU validation set, and the results are presented in Table \ref{tab:mmmu_results}.
We can observe that the scores of vLLM, gLLM, gLLM-epd, and \NAME~are very close, with fluctuations of less than 0.5\%. 
This indicates that \NAME~is capable to maintain functional correctness.

\section{RELATED WORK}

\textbf{Scheduling in LLMs.}
Serving LLMs poses unique scheduling challenges due to variable sequence lengths and heterogeneous computation demands. 
Early systems primarily adopted batch-level scheduling \cite{FasterTransformer}, which is effective for conventional DNN inference but poorly suited for LLM workloads. 
To address this, Orca \cite{Orca} introduced iteration-level scheduling, enabling requests to be admitted or terminated dynamically before full execution. 
However, this design struggles when lengthy prefill requests block subsequent decode requests, resulting in significant latency imbalance. More recently, Sarathi-Serve \cite{Sarathi-serve} proposed chunked prefill, which interleaves prefill and decode by partitioning long sequences into smaller segments. 
Although these approaches improve responsiveness, they overlook multimodal inference and underutilize the parallelism between encoding and prefill. 
To address this, we design \NAME, which unifies intra- and inter-request pipelines to reduce latency and maximize hardware utilization.

\textbf{LLM serving systems.}
To efficiently serve LLMs, several systems have been proposed. Orca \cite{Orca} introduces iteration-level scheduling to improve throughput in distributed serving. For memory efficiency, vLLM~\cite{vllm} employs paged attention to reduce fragmentation, while SGLang~\cite{Sarathi-serve} uses radix attention to eliminate redundant KV cache computations across requests. Splitwise \cite{Splitwise} and DistServe \cite{DistServe} adopt a disaggregated architecture to handle the divergent computational demands of prefill and decode stages by allocating specialized hardware for each phase.
Mooncake \cite{Mooncake}, the serving platform for the Kimi LLM chatbot, further advances disaggregation with a KVCache-centric design, separating prefill and decode clusters while leveraging CPU, DRAM, SSD, and NIC resources to maximize throughput under strict latency SLOs.
Building upon these approaches, \NAME~targets LMM inference, trying to resolve tighter data dependencies between encoding and prefill computation.

\textbf{LMM serving systems.}
Existing research on improving LMM serving can be broadly categorized into algorithm-level inference optimizations and system-level designs. On the algorithm side, Elastic Cache \cite{DBLP:conf/eccv/LiuLWDCRKL24} reduce KV cache overhead through caching and pruning strategies, while Dynamic-LLaVA \cite{DBLP:conf/iclr/HuangZSCZXYL25}, VTW \cite{DBLP:conf/aaai/LinLLJ25}, and QueCC \cite{DBLP:journals/corr/abs-2411-03312} apply token sparsification and compression to vision inputs. These approaches effectively reduce computation and memory costs but often involve efficiency–accuracy trade-offs.
At the system level, ModServe \cite{ModServe} disaggregates multimodal inference pipelines and leverages modality-aware scheduling and autoscaling to handle bursty production traffic with improved throughput and cost efficiency. More recently, another work \cite{EPD} introduces EPD disaggregation, which enables optimizations such as caching multimedia tokens for efficient transfer, parallelizing encoding load, and dynamic role-switching.
Our work, \NAME, complements these directions by focusing on fine-grained scheduling within the inference pipeline. Unlike model-level techniques, \NAME~does not alter model behavior, and unlike ModServe, it directly targets intra- and inter-request pipeline parallelism to reduce latency and improve throughput.

\textbf{Parallelism for LLM training and serving.}
As LLMs continue to grow in size, model parallelism has become indispensable for both distributed training and serving. In training, tensor parallelism, requiring frequent inter-device communication, is widely used in high-bandwidth environments, and recent works \cite{DBLP:conf/asplos/JangdaHLSMMMMS22,DBLP:conf/asplos/WangWSDIHCMMZKG23,DBLP:conf/asplos/ChenLZDSZY24,DBLP:conf/ppopp/DuWJCHCL24} reduce communication idling by overlapping communication with computation. Pipeline parallelism addresses memory imbalance \cite{DBLP:conf/asplos/SunCWF0WC24,DBLP:conf/sc/LiuCZ023,DBLP:conf/icml/KimKYC23}, pipeline bubbles \cite{DBLP:conf/osdi/0075CXPGC0LHCH025,DBLP:conf/sc/LiuCZ023,DBLP:conf/iclr/QiWHL24}, communication overhead \cite{DBLP:conf/ppopp/LinL0WZZ25}, and activation checkpointing \cite{DBLP:conf/asplos/SunCWF0WC24,DBLP:conf/ppopp/LiuLTJ25}. Hybrid strategies combining tensor and pipeline parallelism exploit automated search algorithms \cite{DBLP:conf/osdi/ZhengLZZCHWXZXG22,DBLP:conf/eurosys/ZhangD0CW024,DBLP:conf/usenix/UmOKLKKKMJ24} or heterogeneous hardware characteristics \cite{DBLP:conf/eurosys/ZhangD0CW024,DBLP:conf/usenix/UmOKLKKKMJ24,DBLP:conf/usenix/JiaJWXS0LCLZL022,DBLP:conf/icml/RyabininDDB23}. Frameworks like Megatron-LM \cite{DBLP:conf/sc/NarayananSCLPKV21} provide empirically validated large-scale configurations.
For LLM serving, chunked prefill mechanisms \cite{DBLP:conf/osdi/AgrawalKPMKGTR24} and Token Throttling \cite{gLLM} aim to reduce pipeline bubbles. Mooncake \cite{Mooncake} proposes CPP, which partitions input tokens into chunks processed concurrently across prefill nodes. CPP reduces time-to-first-token, overlaps cross-node communication with computation, and naturally handles both short and long contexts, representing a practical application of pipeline-based acceleration in inference.
The above research on model parallelism optimization is orthogonal to our approach and can serve as a complementary addition to our method.
Combining optimization in tensor parallelism or pipeline parallelism, \NAME~leverages intra- and inter-request parallelism to overlap computation between multimodal encoding and prefill computation.
\section{CONCLUSION}

This paper introduces \NAME, an LMM inference system that efficiently orchestrates both intra- and inter-request pipelines to achieve low latency and high parallelism. 
At the intra-request level, \NAME~leverages a tracker to monitor embedding availability and adopts a stream-style scheduling strategy based on embedding chunk size, enabling fine-grained overlapping between encoding and prefill computations. 
At the inter-request level, \NAME~introduces schedulable tokens to coordinate the execution of multiple requests and fully exploits system parallelism. 
Experimental results on representative LMMs demonstrate that \NAME~reduces latency by up to 66\% and improves throughput by up to 109\%, highlighting its effectiveness in accelerating LMM inference.

\bibliographystyle{ACM-Reference-Format}
\bibliography{ref}

\end{document}